\documentclass[11pt]{article}
\usepackage[a4paper,left=3cm,right=3cm]{geometry}
\usepackage[noblocks]{authblk}
\usepackage{amsmath}
\usepackage{amssymb}
\usepackage{amsthm}
\usepackage{bm}
\usepackage{breqn}
\usepackage{framed}
\usepackage[table]{xcolor}
\usepackage{graphicx}
\usepackage{subfigure}
\usepackage{multirow}

\colorlet{light}{gray!20}
\colorlet{dark}{gray!35!white}

\newtheorem{theorem}{Theorem}

\newcommand{\EX}{\mathbb{E}}

\begin{document}
\title{Realization of Waddington's Metaphor: Potential Landscape, Quasi-potential, A-type Integral and Beyond}
\author{Peijie Zhou\thanks{E-mail: {\tt cliffzhou@pku.edu.cn}}}
\author{Tiejun Li\thanks{E-mail: {\tt tieli@pku.edu.cn}. Corresponding author. }}
\affil{LMAM and School of Mathematical Sciences, Peking University, Beijing 100871, China}
\maketitle

\begin{abstract}
Motivated by the famous Waddington's epigenetic landscape metaphor in developmental biology, biophysicists and applied mathematicians made different proposals to realize this metaphor in a rationalized way. We adopt comprehensive perspectives to systematically investigate three different but closely related realizations in recent literature: namely the potential landscape theory from the steady state distribution of stochastic differential equations (SDEs), the quasi-potential from the large deviation theory, and the construction through SDE decomposition and A-type integral.The connections among these theories are established in this paper. We demonstrate that the quasi-potential is the zero noise limit of the potential landscape. We also show that the potential function in the third  proposal coincides with the quasi-potential. The most probable transition path by minimizing the Onsager-Machlup or Freidlin-Wentzell action functional is discussed as well. Furthermore, we compare the difference between local and global quasi-potential through the exchange of limit order for time and noise amplitude.  As a consequence of such explorations, we arrive at the existence result for the SDE decomposition while deny its uniqueness in general cases. It is also clarified that the A-type integral is more appropriate to be applied to the decomposed SDEs rather than the original one. Our results contribute to a better understanding of existing landscape theories for biological systems.
\end{abstract}

\section{Introduction: Landscape Theories for Biological Systems}

Published in 1957,  the Waddington's epigenetic landscape metaphor \cite{waddington1957strategy} provides a vivid pictorial description as well as an insightful qualitative tool to understand the mechanism of gene regulation in evolutionary and developmental biology \cite{jamniczky2010rediscovering}. In recent years, we witness the growing interests and efforts to quantitatively realize this metaphor in a rationalized way and utilize the constructed energy landscape to study the robustness, adaptivity and efficiency of biological networks. In this paper, we will focus on three representative and closely related works among them:   (i) The potential landscape theory from the steady state distribution of stochastic differential equations (SDEs); (ii) the quasi-potential from the large deviation theory (LDT) and (iii) Ao's construction through SDE decomposition and  A-type integral (it will be called {\it SDE decomposition theory} below for short). To clarify the connection and difference among them is the main concern of this paper.

The considered three theories were proposed from different motivations and backgrounds. Enlightened by the Boltzmann distribution law in equilibrium statistical mechanics,  J. Wang et al. \cite{wang2008potential} constructed the potential landscape from the steady-state distribution of non-equilibrium biological systems and adopted it in the analysis of many real biological models including budding yeast cell cycle \cite{wang2010potential}, stem cell differentiation \cite{wang2011quantifying} and Calcium oscillation \cite{xu2013potential}, etc. Arising in Freidlin and Wentzell's study on LDT  for diffusion processes \cite{freidlin2012random}, the quasi-potential was proposed by minimizing the LDT rate functional between different states and has been applied in genetic switching models \cite{lv2014constructing,chen2015distinguishing} and cell cycle dynamics \cite{lv2015energy}. Motivated by the fluctuation-dissipation theorem, P. Ao and his coworkers performed the SDE decomposition to obtain the underlying potential function \cite{zhu2004robustness,ao2004potential} and proposed the so-called A-type integral interpretation of the SDEs \cite{yuan2012beyond}.

Although these existing theories on energy landscape yield fruitful applications in real biological systems,  to the authors' knowledge, very limited work has been done to elucidate their relationships and connections. In \cite{zhou2012quasi}, an overview related to these theories was presented for biological systems under extrinsic perturbations. In this paper, we will continue the discussion in a more general setup by considering the diffusion process of the form
\begin{equation}\label{eqn:original}
dX_{t}=b(X_{t})dt+\sigma(X_{t})dW_{t},\quad \sigma(x)\sigma(x)^{t}=2\varepsilon D(x),
\end{equation}
where $X_{t}, b(X_{t})\in\mathbb{R}^{n}, W_{t}\in\mathbb{R}^{m}$ and $\sigma(x)\in\mathbb{R}^{n\times m}$. The subscript $t$ means the time dependence instead of time derivative. $W_{t}$ is the standard Brownian motion with mean $\EX W_{t}=0$ and covariance $\EX (W_{s}W_{t})=\min(s,t)$. $\varepsilon$ represents the strength of noise.  Here we employ the notation $dW_{t}$ as in probability theory since $\dot{W}_{t}$ is not an ordinary function mathematically \cite{Gardiner2004Book}.  Unless otherwise stated, the stochastic integral is understood in Ito sense.
The SDEs (\ref{eqn:original}) is an abstraction of chemical Langevin equation \cite{gillespie2000chemical} to model chemical reactions in the biological systems with $n$ species of reactants and $m$ reaction channels, where the random vector $X_{t}$ denotes the concentration of different chemicals in the reaction network at time $t$. The state-dependent diffusion matrix $D(x)$ enables us to investigate the intrinsic noise of chemical reactions, which is an inherent property of biological networks \cite{elowitz2002stochastic}. It also includes the extrinsic noise as a special case.   Since Waddington's metaphor describes the cell development as the motion of marbles among the valleys, it is also helpful to interpret $b(x)$ as the ``force'' and $D(x)$ as the ``diffusion coefficient" in over-damped Langevin dynamics.

The purpose of this paper is two-fold. Firstly, we will reveal the relationships among the three landscape theories for biological systems modeled by SDEs (\ref{eqn:original}). Though different theories may be proposed from different specific perspectives, analyzing their mathematical structure could establish the connection among them. We will show that the potential landscape is consistent with the quasi-potential when the noise amplitude tends to zero, and the quasi-potential exactly coincides with the potential function in SDE decomposition theory under certain conditions. Secondly, as a by-product of the established connections, we get some new insights and findings on the existing landscape theories. Specifically we will provide a mathematically rigorous existence result for SDE decomposition theory, and show that under the current framework of the proposal \cite{ao2004potential}, the decomposition is generally not unique when the dimension of SDEs (\ref{eqn:original}) is bigger than or equal to 3. As a corollary, we clarify that the A-type integral interpretation for SDEs (\ref{eqn:original}) might be ill-defined and thus only be applied to a known decomposed form.

The paper is organized as follows. In Section 2, we study the potential landscape theory defined from the steady state distribution perspective and explore its physical and biological meaning through force decomposition. The difficulties of studying transition path under the potential landscape framework naturally leads to the discussion in Section 3 on quasi-potential within Freidlin and Wentzell's framework. The consistency between potential landscape and global quasi-potential in the small noise limit,  and the connection between local and global quasi-potential will be investigated there. In Section 4, we review the SDE decomposition theory and show the coincidence between the decomposed potential function and the quasi-potential. The existence and non-uniqueness of the SDE decomposition is also presented. In Section 5, an example of constructing energy landscape for the diffusion on the circle $\mathbb{S}[0,1]$ through different theories is provided as further explanation. Finally we discuss the implications of our results and some future topics  in Section 6.

\section{Potential Landscape Theory}

The potential landscape theory proposed by J. Wang et al. is defined through the steady state distribution, which is a generalization of Boltzmann's distribution law in equilibrium statistical mechanics. It has been widely applied to many biological systems and become an important methodology \cite{Wang2015Review}. While the biological systems are not in equilibrium in general,  this non-equilibrium feature naturally appears if we consider the force decomposition using the derived potential landscape function.  Associated with the potential landscape theory with finite noise, the optimal transition path is obtained by minimizing the Onsager-Machlup functional, which raises challenge when the admissible transition time tends to  infinity. This issue will be resolved in the quasi-potential theory in the zero noise limit.

\subsection{Starting Point: Steady State Distribution}
The starting point of constructing Wang's potential landscape is the steady state distribution of SDEs (\ref{eqn:original}). We have the Fokker-Planck equation of SDEs (\ref{eqn:original}) 
\begin{equation}\label{eqn:fpe}
\partial_{t} P(x,t)+\nabla\cdot J(x,t)=0,
\end{equation}
where $P(x,t)$ is the probability distribution density (PDF) of the process $X_{t}$ at time $t$, and the probability flux
\begin{equation*}
J(x,t)=b(x)P(x,t)-\varepsilon\nabla\cdot (D(x)P(x,t)).
\end{equation*}
 The steady state distribution $P_{ss}(x)$ and steady state probability flux $J_{ss}(x)$ can be obtained by solving
\begin{equation}
\nabla\cdot J_{ss}(x)=\nabla\cdot [b(x)P_{ss}(x)-\varepsilon\nabla\cdot (D(x)P_{ss}(x))]=0.
\label{eqn:steady state}
\end{equation}
Then the potential landscape function $\phi^{PL}(x)$ is defined as
\begin{equation}\label{eqn:pl}
\phi^{PL}(x)=-\ln P_{ss}(x).
\end{equation}
The relationship $P_{ss}(x)=\exp(-\phi^{PL}(x))$ is reminiscent of the Boltzmann-Gibbs distribution in equilibrium statistical mechanics. 

The rationale of the potential landscape can be shown explicitly if we consider a gradient system with
$$b(x)=-\nabla V(x),~D(x)=I.$$
We have 
$$P_{ss}(x) = \frac{1}{Z}\exp\Big(-\frac{1}{\varepsilon}V(x)\Big),\quad J_{ss}(x)=0,$$
thus
\begin{equation}\label{eqn:PLC}
\phi^{PL}(x)=\frac{1}{\varepsilon}V(x)+\ln Z.
\end{equation}
In this case, the potential landscape $\phi^{PL}(x)$ is equivalent to the original driving potential $V(x)$ up to a rescaling and a shift. But of course, this observation does not hold for general $b(x)$ or $D(x)\neq I$, in which case one gets a generalized potential.

In practice, there are mainly two approaches to numerically compute $P_{ss}(x)$ and therefore $\phi^{PL}(x)$. The most direct approach is to solve the Fokker-Planck Eq.  (\ref{eqn:fpe}) by applying deterministic numerical methods with appropriate boundary condition.  However, the computational cost of such strategy increases exponentially, and quickly becomes unaffordable even when the dimension $n\ge 4$.  Hence in high dimensional cases, $P_{ss}(x)$ is either obtained by exploring the special feature of the considered dynamics, e.g. the mean field approximation \cite{wang2010potential}; or obtained by direct Monte-Carlo simulation of SDEs (\ref{eqn:original}) until steady state distribution. However, this approach also encounters the difficulty of slow convergence when the noise strength $\varepsilon$ is very small, in which case the metastability and ergodicity turn to be key issues \cite{Frenkel2002Book}. Moreover, both the representation and storage of the high dimensional potential landscape need to be studied at first. We confront with the curse of dimensionality.

\subsection{Force Decomposition: Non-Equilibrium Steady States}\label{sec:ForceDecNESS}

Wang's potential landscape theory can also be studied from force decomposition perspective. From the relationship between flux and probability
\begin{equation*}
J_{ss}(x)=b(x)P_{ss}(x)-\varepsilon\nabla\cdot (D(x)P_{ss}(x)),
\end{equation*}
and Eq. (\ref{eqn:pl}), we can represent the drift term $b(x)$ in the decomposition form
\begin{equation}\label{eqn:wangdecom}
\begin{split}
b(x) &=\frac{\varepsilon}{P_{ss}(x)}\nabla\cdot (D(x)P_{ss}(x))+\frac{J_{ss}(x)}{P_{ss}(x)}\\
       &= -\varepsilon D(x)\nabla\phi^{PL}(x)+\varepsilon\nabla\cdot D(x)+\frac{J_{ss}(x)}{P_{ss}(x)}.
\end{split}
\end{equation}
To gain more intuitions from (\ref{eqn:wangdecom}), let us specifically take $D(x)=I$ and $\varepsilon=1$. Then
\begin{equation*}
b(x) = -\nabla\phi^{PL}(x)+\frac{J_{ss}(x)}{P_{ss}(x)}.
\end{equation*}
We will discuss two cases for different values of $J_{ss}(x)$ and their biological meaning. 

The first case is  $J_{ss}(x)=0$. Such condition is called the {\it detailed balance} in probability theory, while in statistical mechanics,  systems with zero flux correspond to the equilibrium states.  Under such circumstances, the force is simply the negative gradient of the potential landscape, i.e. $b(x) = -\nabla\phi^{PL}(x)$. Hence viewing from the biological perspective, the detailed balance condition implies the equilibrium states where the biological system is driven by the gradient of potential landscape and the steady state distribution is of the Boltzmann-Gibbs form $P_{ss}(x)=\exp(-\phi^{PL}(x))$.

The second case is $J_{ss}(x)\not=0$, which is more common in biological systems. Under such circumstances, when the system reaches steady state, the probability flux does not vanish, leading to the non-equilibrium steady states (NESS). The force $b(x)$ is now decomposed into the gradient term $-\nabla\phi^{PL}(x)$ and an additional non-gradient term $J_{ss}(x)/P_{ss}(x)$, which is also called ``curl" term because $\nabla\cdot J_{ss}(x)=0$. One typical example of NESS in biological models is the oscillatory dynamics, because the limit cycle can not exist in gradient systems and must be driven by the curl term. Many concepts in non-equilibrium statistical mechanics such as entropy production can be analyzed within NESS framework and one may consult \cite{zhang2012stochastic} for a systematic survey. 

A simple illustrative example to show the construction of the $\phi^{PL}(x)$ and the non-equilibrium nature of NESS will be presented in Section \ref{sec:example}. 

\subsection{Transition Path: Path Integral and Challenges}

In Waddington's metaphor, the concept of transition path corresponds to the switching of biological systems among different meta-stable states. Below we will mainly focus on establishing the connections between the potential landscape and the transition path through path integral formulation. 

Following Feynman's path integral approach to quantum mechanics \cite{feynman1948space}, we can also solve the Fokker-Planck Eq.  (\ref{eqn:fpe}) formally by integrating the individual paths $\psi(t)$ according to their weight \cite{wang2010kinetic}
\begin{equation}\label{eq:PathInt}
\mathbb{P}(x_{f},T| x_{i}, 0)=\int \mathcal{D}\psi P(\psi|\psi(0)=x_{i},\psi(T)=x_{f}) =\frac{1}{Z}\int \mathcal{D}\psi \exp(-S_{T}[\psi]),
\end{equation}
where $x_{f}$ denotes the final state, $x_{i}$ denotes the initial state and $Z$ is the partition function in path space. The weight of each path $P(\psi|\psi(0)=x_{i},\psi(T)=x_{f})$ is assigned according to its action functional $S[\psi]$. For diffusion process, the action functional can be expressed as the time integral of Onsager-Machlup Lagrangian function \cite{fujita1982onsager}
\begin{equation}
S_{T}[\psi]=\int_{0}^{T}L^{OM}(\psi,\dot{\psi})ds,
\label{eqn:omaction}
\end{equation}
whose concrete form will be discussed later. If the SDEs (\ref{eqn:original}) is ergodic and suppose the system starts from $x_{0}$, then $P_{ss}(x)=\lim\limits_{T\to\infty}\mathbb{P}(x,T| x_{0}, 0)$, yielding the {\it formal} relationship between potential landscape and transition path
\begin{equation}\label{eqn:pltp}
\phi^{PL}(x)=-\ln P_{ss}(x)=-\lim\limits_{ T\to +\infty}[\ln\int \mathcal{D}\psi \exp\left(-\int_{0}^{T}L^{OM}(\psi,\dot{\psi})ds\right)-\ln Z].
\end{equation} 

Several problems will be encountered in this formal treatment of potential landscape theory from transition path perspective.

Firstly, if we want to compute the potential landscape at point $x$ from transition path perspective, one may have to sum up the weights over all transition paths starting from one given point $x_{0}$ and reaching $x$ as time goes to infinity. Constructing numerical algorithms to compute potential landscape directly from such tactics turns out to be a challenging task. 

To avoid such inconvenience, it is developed in \cite{wang2010kinetic} that another version of landscape function can be constructed from the ``effective action'' of the ``dominant path''. To briefly state, the landscape function at point $x$ equals to the minimum action $S_{T}[\psi]$ in Eq. (\ref{eqn:omaction}) of all the paths $\psi$ connecting  $x$ to the reference point $x_{0}$ (usually the attractor of the system) in this theoretical framework. The minimum action path is dominant especially when $\varepsilon$ is small because it corresponds to the maximum weight path in Eq. (\ref{eq:PathInt}). However, such proposed landscape function might not be well-defined for some systems. For instance, let us consider a specific gradient system with 
$$b(x)=-\nabla V(x),~D(x)=I, ~\varepsilon=1,~V(x)=\frac{1}{2}x^{2}.$$
Take $x_{0}=0$, then the action functional now has the concrete form (see the discussion below for general cases)  
$$S_{T}[\psi]=\int_{0}^{T}L^{OM}(\psi,\dot{\psi})ds=\int_{0}^{T}\frac{1}{4}(\dot{\psi}+\psi)^2 ds-\frac{1}{2}T,~\psi(0)=x,  ~\psi(T)=0.$$
Hence the minimum action path $\hat{\psi}$ satisfies $d\hat{\psi}/dt=-\hat\psi,\hat{\psi}(0)=x,\hat{\psi}(+\infty)=0$, also indicating that $S_{T}(\hat{\psi})=-\infty$.
In this case, we have that the landscape function proposed in \cite{wang2010kinetic} at every point $x\not=0$ is minus infinity, which is not a desirable result. We remark that this phenomenon actually results from the divergence term $\frac{1}{2}\nabla\cdot b(x)$ in the OM function.     

Moreover, the choice of concrete OM function form $L^{OM}(\psi,\dot{\psi})$ for the general diffusion process is a rather subtle and controversial issue. It is shown in \cite{durr1978onsager} that if the diffusion matrix $D(x)$ is constant and $n=1$, the most probable path (i.e. the path with largest weight) correspond to the minimizer of action function with Lagrangian 
\begin{equation}\label{eqn:OM}
L^{OM}(\psi,\dot{\psi})=\frac{1}{4\varepsilon}[\dot{\psi}-b(\psi)]^{t}D^{-1}(\psi)[\dot{\psi}-b(\psi)]+\frac{1}{2}\nabla\cdot b(\psi).
\end{equation}
For the state-dependent diffusion matrix $D(x)$, it is argued in \cite{feng2014non} that the term $\frac{1}{2}\nabla\cdot b(\psi)$ in (\ref{eqn:OM}) should be replaced by $\sum\limits_{i,j,k}\frac{1}{2}D_{ij}\partial_{x_{j}}(D^{-1}_{ik}b_{k})$. While in \cite{tang2014summing}, it is claimed that an additional term involving second order derivative of $\sigma(x)$ (only the one dimensional case is considered in \cite{tang2014summing}) should be added in order to give the correct solution for Ito-type Fokker-Planck Equation. We remark that the general mathematical expression for OM function in high dimensional cases has been studied in \cite{takahashi1981probability,capitaine2000onsager}, which include the result in \cite{tang2014summing} as a special case.

However, the above difficulties arising in the study of potential landscape from the transition path perspective can be all resolved under the regimes of small noise limit. If $\varepsilon$ is sufficiently small, then only the term $\frac{1}{4\varepsilon}[\dot{\psi}-b(\psi)]^{t}D^{-1}(\psi)[\dot{\psi}-b(\psi)]$ will count in 
OM function (\ref{eqn:OM}), which corresponds to the Friedlin-Wentzell (FW) function whose form is widely acknowledged and accepted. Moreover, the weight of the most probable path will dominate in (\ref{eqn:pltp}) according to Laplace's integral asymptotics as $\varepsilon$ tends to zero, indicating that we can use the minimum action based on FW function, which is well-defined,  to describe the landscape instead of summing up the weights over all transition path. This observation leads to the introduction of quasi-potential below, whose theory has been well established within the rigorous mathematical framework of Freidlin and Wentzell's Large Deviation Theory (LDT) for diffusion processes \cite{freidlin2012random}.

A summary for the discussions above on the potential landscape theory is provided in Table \ref{table:pl}.

\begin{table}[h]
\centering
\rowcolors{1}{dark}{light}
\renewcommand\arraystretch{2}
\begin{tabular}{p{5cm} p{9 cm}}
\multicolumn{2}{c}{\bf Realization of Waddington's Metaphor Candidate I: Potential Landscape} \\

\bf Definition & $\phi^{PL}(x)=-\ln P_{ss}(x)$ (Steady State Distribution) \\

& Deterministic numerical method (e.g. difference method) for Fokker-Planck Equation (in low dimensional system) \\
\rowcolor{dark}\multirow{-2}{*}{\bf Numerical Strategy} & Monte-Carlo simulation for Stochastic Differential Equation (in high dimensional system, inefficient when noise strength $\varepsilon$ is small)\\

\rowcolor{light}\bf Force Decomposition & $b(x)= -\varepsilon D(x)\nabla\phi^{PL}(x)+\varepsilon\nabla\cdot D(x)+J_{ss}(x)/P_{ss}(x)$, where $J_{ss}(x)/P_{ss}(x)$ reflects the NESS nature of the system \\

\rowcolor{dark}\bf Transition Path Perspective & $\phi^{PL}(x)=-\lim\limits_{T\to\infty}\ln\frac{1}{Z}\int \mathcal{D}\psi \exp(-\int_{0}^{T}L^{OM}(\psi,\dot{\psi})ds)$, where the integral is over all the paths $\psi$ satisfying $\psi(T)=x$
\end{tabular}

\caption{Summary for the potential landscape $\phi^{PL}(x)$.}
\label{table:pl}
\end{table}
\section{Quasi-Potential Theory}

As has been discussed above, the quasi-potential theory aims to quantify the landscape for biological system whose noise amplitude $\varepsilon$ is small enough, which is a reasonable assumption when the number of molecules is large. Although the quasi-potential theory has rigorous mathematical formulation, we will continue adopting the path integral formulation to see how the small noise assumption can help simplify the treatment of transition path discussed above. When establishing the connections between potential landscape and quasi-potential, the concept of global quasi-potential $\phi^{QP}(x)$ will be naturally brought out. They can be connected via the WKB asymptotics applied to the steady state of the Fokker-Planck equation. This leads to the steady  Hamilton-Jacobi equation for the quasi-potential  $\phi^{QP}(x)$ and another type of decomposition of the force $b(x)$.

\subsection{Starting Point: Definition and Transition Path }

Let $X^{\varepsilon}_{t}$ denote the trajectory of SDE (\ref{eqn:original}). The Freidlin-Wentzell theory roughly tells that for a given regular connecting path $\psi(t)$ and $\varepsilon$, $\delta$ small enough,  we have
\begin{equation}
\mathbb{P}(\sup_{0\leq t\leq T} |X^{\varepsilon}_{t}-\psi(t)|\leq\delta)\approx \exp(-\varepsilon^{-1}S_{T}[\psi]).
\label{eqn:ldt}
\end{equation}
The action functional $S_{T}[\psi]$ is also  called the {\it rate functional} in LDT with the expression
\begin{equation*}
S_{T}[\psi]=
\begin{cases}
\int_{0}^{T} L^{FW}(\psi,\dot{\psi})dt,\quad \text{if $\psi$ absolutely continuous and integral converges,}\\
+\infty,\quad\text{otherwise,}
\end{cases}
\end{equation*}
where 
\begin{equation*}
L^{FW}(\psi,\dot{\psi})=\frac{1}{4}[\dot{\psi}(s)-b(\psi(s))]^{t}D^{-1}(\psi(s))[\dot{\psi}(s)-b(\psi(s))]
\end{equation*}
is the dominate $O(\varepsilon^{-1})$ term in the Onsager-Machlup functional. We also call $S_{T}[\psi]$ the Freidlin-Wentzell functional in later text. The approximation \eqref{eqn:ldt} is indeed derived by applying Laplace asymptotics to the path integral formulation (c.f. Appendix A). Borrowing the idea from classical mechanics, we call $L^{FW}(\psi,\dot{\psi})$ the Lagrangian of action $S_{T}$, and correspondingly  define the Hamiltonian of the system by taking the Legendre dual of the Lagrangian \cite{landau1960classical}
\begin{equation}\label{eq:Hamilton}
H(\psi,p)=b(\psi)^{t} p+p^{t}D(\psi)p.
\end{equation}

Assume $x_{0}$ is a stable fixed point of the deterministic dynamical system $dx/dt=b(x)$, representing a meta-stable biological state. Then the {\it local} quasi-potential at state $x$ with respect to $x_{0}$ is defined as
\begin{equation}\label{eqn:qp}
\phi^{QP}_{loc}(x;x_{0})=\inf\limits_{T>0}\inf\limits_{\psi(0)=x_{0},\psi(T)=x}\int_{0}^{T}L^{FW}(\psi,\dot{\psi})dt.
\end{equation}
The heuristic explanation of this definition is that the energy difference  between state $x$ and $x_{0}$ can be evaluated by the least action cost of moving the system from $x_{0}$ to $x$, because only the minimum action path dominates in Eq. (\ref{eqn:pltp}) in the limit $\varepsilon\rightarrow 0$.

To understand the intuition behind the quasi-potential, let us consider a gradient dynamics with a single-well potential $V(x)$, i.e.
$$b(x)=-\nabla V(x),~D(x)=I.$$
We assume that $V(x)\geq 0$ and $V(x_0) = 0$
is the unique minimum of $V(x)$. By definition, we have
\begin{equation}
  \label{eq:SLg}
   S_T[\psi] = \frac{1}{4}\int_0^T |\dot{\psi}+\nabla V(\psi)|^2dt.
\end{equation}
First we show that $S_T[\psi]\geq V(x)$ for all $\psi$ with endpoints $\psi(0) = x_0$ and $\psi(T) = x$ because 
\begin{align}
  S_T[\psi] &= \frac{1}{4}\int_0^T |\dot{\psi}+\nabla V(\psi)|^2dt\nonumber \\
  &= \frac{1}{4}\int_0^T |\dot{\psi}-\nabla V(\psi)|^2dt  +
  \int_0^T \dot{\psi}^{t}\nabla V(\psi)dt\nonumber \\
  &\geq \int_0^T \dot{\psi}^{t}\nabla V(\psi)dt=\int_{0}^{T} dV(\psi) = V(x).
\end{align}
On the other hand, we can choose a special $\hat{\psi}$ and $T>0$ such that $\dot{\hat{\psi}} = \nabla V(\hat{\psi})$ and $\hat\psi(T)=x$ (this $T$ equals $\infty$ indeed). For
this special $\hat{\psi}$,
\begin{equation}
S_T[\hat{\psi}] =  \int_0^T \dot{\hat\psi}^{t}\nabla V(\hat\psi)dt = V(x).
\end{equation}
The above discussion shows that $\phi^{QP}_{loc}(x;x_0) = V(x)$ in this single-well gradient case. The quasi-potential generalizes the potential concept in general situation.

The LDT result \eqref{eqn:ldt} also implies that the minimizer of the variational problem \eqref{eqn:qp} gives the minimum action path or most probable path connecting two metastable states in zero noise limit. It can be shown \cite{freidlin2012random,Arnold1989Book} that the minimizer $\psi(t)$ satisfies Hamilton's canonical equation 
\begin{equation}
\dot{\psi}(t)=\nabla_{p} H(\psi,\nabla_{x}\phi^{QP}_{loc}(\psi;x_{0}))=b(\psi)+2D(\psi)\nabla_{x}\phi^{QP}_{loc}(\psi;x_{0}),\quad \psi(0)=x_{0}.
\end{equation}
From the fact $\phi^{QP}_{loc}(x_{0};x_{0})=0$ and $\phi^{QP}_{loc}(x;x_{0})\ge 0$, one has that
$x_{0}$ is the global minimum of $\phi^{QP}_{loc}(x;x_{0})$ and thus $\nabla_{x}\phi^{QP}_{loc}(x_{0};x_{0})=0$. Combing with the condition $b(x_{0})=0$ we know that the minimum action path must reach $x$ in infinite time in Eq. (\ref{eqn:qp}).

To compute the local quasi-potential and minimum action path numerically, one possible strategy is to derive and solve the Euler-Lagrange equation of variational problem (\ref{eqn:qp}). However, we will generally encounter a singular boundary value problem  because the system does not reach $x$ in finite time. This difficulty can be overcome by applying the geometric minimum action method (gMAM) to solve the variational problem (\ref{eqn:qp}) directly through Maupertuis principle in the space of curves \cite{vanden2008geometric, lv2014constructing, lv2015energy}.

\subsection{Steady State Distribution: Local and Global Quasi-Potential}

To connect the quasi-potential and potential landscape, we need to consider the {\it global} quasi-potential. The relation between the local and global version of quasi-potential can be understood from the exchange of limit order for noise strength $\varepsilon$ and time $t$, which is also meaningful in biology (c.f. Section \ref{sec:DiscussExch}). Another connection is by noting that both are solutions of a specific steady Hamilton-Jacobi equation to be shown in Section \ref{sec:HJE}.

\subsubsection{Local and Global Quasi-potential}\label{sec:localqp}

Viewing from steady state distribution perspective, the limit that we desire to compute is 
\begin{equation}\label{eqn:globallim}
\lim_{\varepsilon \to 0}-\varepsilon\ln P_{ss}(x)=-\lim_{\varepsilon \to 0}\lim_{t \to +\infty}\varepsilon\ln P^{\varepsilon}(x,t|x_{0},0),
\end{equation}
where $P^{\varepsilon}(x,t|x_{0},0)$ denotes the transition PDF from one stable fixed point $x_{0}$ at $t=0$ to state $x$ at time $t$. Here we assume the system is ergodic and the stochastic dynamics starts from $x_{0}$. 

Adopt the path integral interpretation \eqref{eq:PathInt} and apply Laplace's method in path space formally (c.f. Appendix A), we get
\begin{equation}
\begin{split}
-\lim_{\varepsilon \to 0}\varepsilon\ln P^{\varepsilon}(x,t|x_{0},0) &=-\lim_{\varepsilon \to 0}\varepsilon\Big[\ln \int \mathcal{D}\psi \exp(-\varepsilon^{-1}S_{t}[\psi]) -\ln Z\Big]\\
&=\inf\limits_{\psi(0)=x_{0},\psi(t)=x}S_{t}[\psi], \\
\end{split}
\end{equation}
where $S_{t}[\psi]$ corresponds to the Freidlin-Wentzell functional since the higher order terms disappear in the zero noise limit. Correspondingly we obtain
\begin{equation}\label{eqn:locallim}
\begin{split}
-\lim_{t \to +\infty}\lim_{\varepsilon \to 0}\varepsilon\ln P^{\varepsilon}(x,t|x_{0},0) &=\lim_{t \to +\infty}\inf\limits_{\psi(0)=x_{0},\psi(t)=x}S_{t}[\psi]\\
&=\phi^{QP}_{loc}(x;x_{0}).
\end{split}
\end{equation}
Although the above equations are formally established through the path integral approach, whose rigorous mathematical rationality needs to be further explored, we can gain some heuristic findings from such treatments.
We observe that the difference between the left hand side (LHS) of Eq. (\ref{eqn:locallim}) and the right hand side (RHS) of Eq.  (\ref{eqn:globallim}) is just the exchange of limit order for $t$ and $\varepsilon$. Intuitively we would claim that the exchange of limit order yields different results,  because $P_{ss}(x)$ is a global quantity for the stochastic dynamics while $\phi^{QP}_{loc}(x;x_{0})$ only reflects the local information with respect to a specific stable state $x_{0}$. In Freidlin-Wentzell theory, the limit of Eq.  \eqref{eqn:globallim} is called the global quasi-potential  $\phi^{QP}_{glob}(x)$, i.e.
\begin{equation}\label{eqn:globqp}
\lim_{\varepsilon \to 0}-\varepsilon\ln P_{ss}(x)=-\lim_{\varepsilon \to 0}\lim_{t \to +\infty}\varepsilon\ln P^{\varepsilon}(x,t|x_{0},0)=\phi^{QP}_{glob}(x).
\end{equation}

The distinction between the limits in (\ref{eqn:locallim}) and \eqref{eqn:globqp} can be understood from the separation of time scales. For a stochastic dynamical system with multiple attractors, the system exhibits different behavior in different time scales.  In short time scale $\tau_{S}$ the system will walk around one specific attractor, while in longer time scale $\tau_{L}$ the system will transit among different attractors. According to LDT or Kramers' theory, the scale separation between long and short time scales is of order $\tau_{L}/\tau_{S}=\exp(\Delta V/\varepsilon)$, where $\Delta V$ represents the characteristic barrier height between different attractors. In the limit order in  (\ref{eqn:locallim}), the large enough time $t$ is fixed first, and $\varepsilon$ can be chosen sufficiently small such that $t\sim O(\tau_{S})$ with respect to $\varepsilon$. Hence the limit  $\phi^{QP}_{loc}(x;x_{0})$ only reflects the local information about $x_{0}$ because the system mainly fluctuates around the stable point and could not see the outside region in this regime.  In comparison, when the limit order in (\ref{eqn:globallim}) is considered, the small noise $\varepsilon$  is fixed first, and we can wait sufficiently long time such that $t\sim O(\tau_{L})$ with respect to $\varepsilon$. Therefore the limit $\phi^{QP}_{loc}(x;x_{0})$ can tell about the global behavior of the system because transitions among different states are common under such circumstance. 

In biological systems, the noise exists however small it is, and the amplitude of intrinsic noise is an inherent nature of the system determined by system size.  Hence when studying the long time behavior of certain system (e.g. cell differentiation), it is more appropriate to view the noise amplitude as fixed a priori  while the time as dependent on the observation. In this sense, the limit order in Eq.  (\ref{eqn:globallim}) is more realistic for biological systems, which suggests that the global quasi-potential is a more advisable candidate to quantify the Waddington's metaphor rather than the local version. For simplicity, we will just call the global quasi-potential $\phi^{QP}_{glob}(x)$ as quasi-potential $\phi^{QP}(x)$ in later text. From (\ref{eqn:globqp}) we reach the relationship between potential landscape and quasi-potential 
\begin{equation}
\lim_{\varepsilon \to 0} \varepsilon\phi^{PL}(x)=\phi^{QP}(x).
\label{eqn:ssqp}
\end{equation}
This fact can be also observed from \eqref{eqn:PLC} as a special case.

\subsubsection{Constructing Global Quasi-potential From Local Ones}

Unexpectedly, the global quasi-potential can be constructed from local ones with an interesting sticking procedure \cite{freidlin2012random}. We will use a simple example to illustrate this point.

\begin{figure}[htbp]
\centering 

\subfigure[The schematics of double-well potential $V(x)$.]
{
\label{fig:potential}
\includegraphics[width=0.45\textwidth]{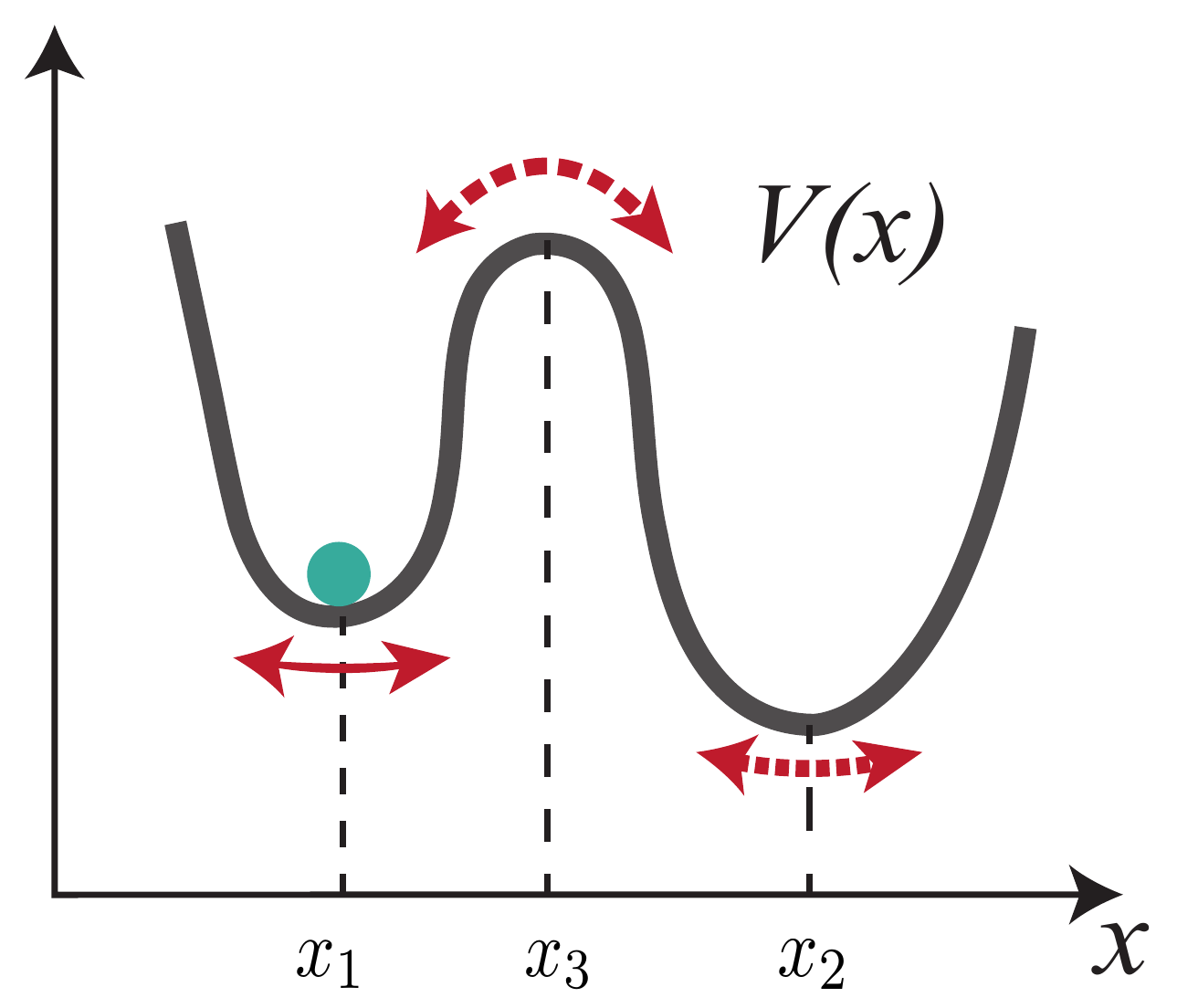}
}
\hspace*{5mm}
\subfigure[Local quasi-potential constructed from $x_1$.]
{
\includegraphics[width=0.44\textwidth]{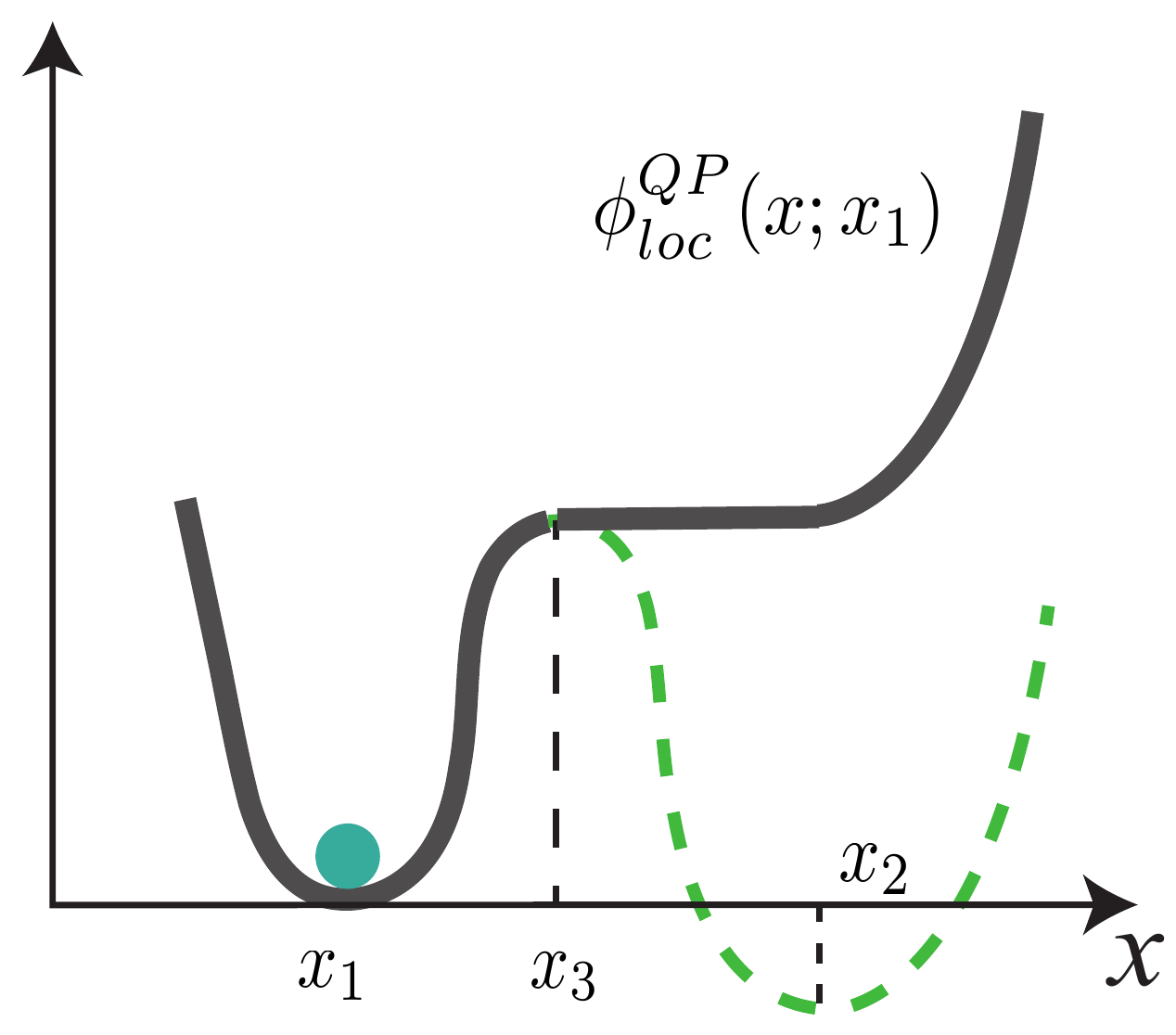}
}
\subfigure[Local quasi-potential constructed from $x_2$.]
{
\includegraphics[width=0.45\textwidth]{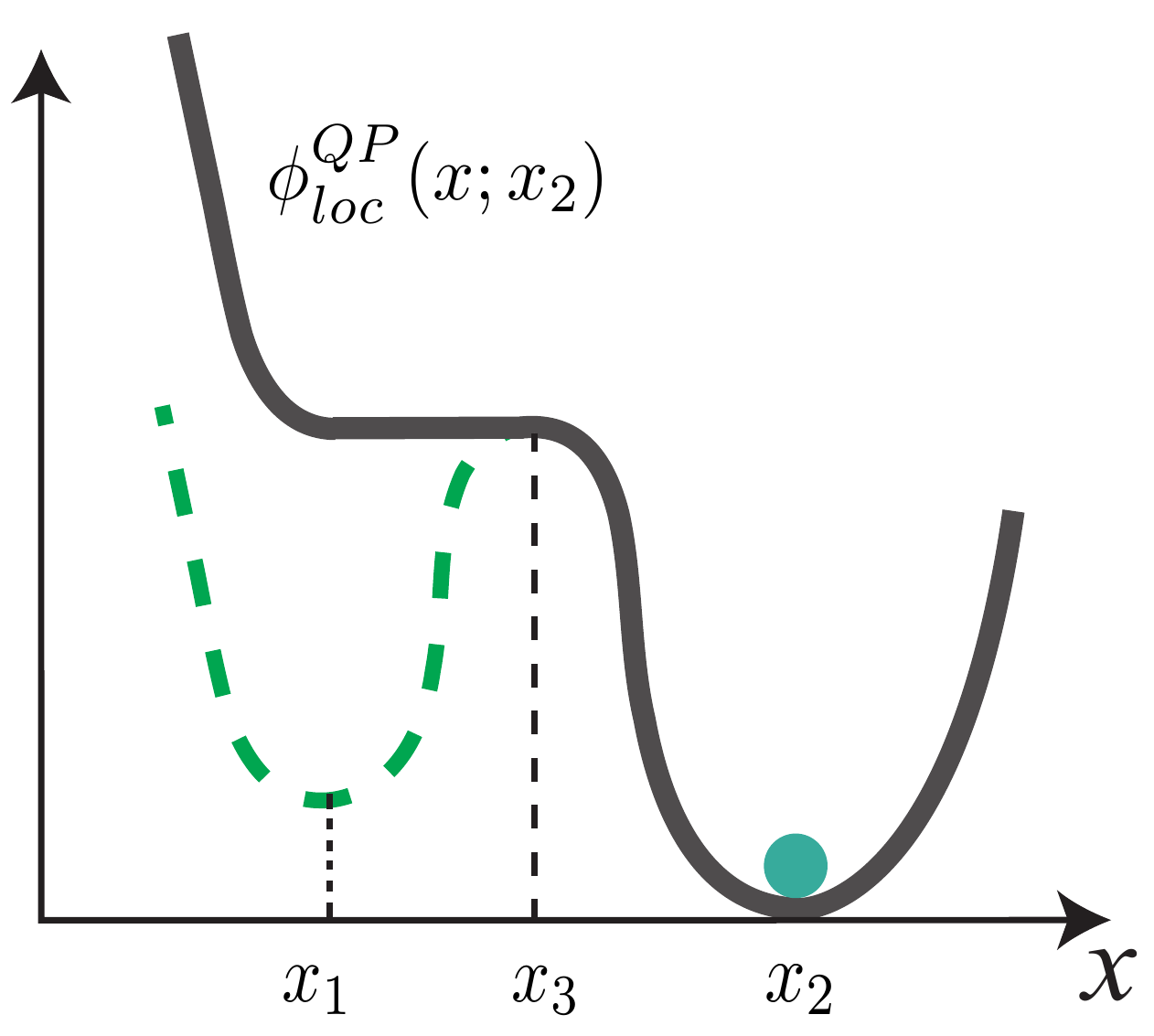}
}
\hspace*{5mm}
\subfigure[The constructed global quasi-potential is a shift of $V(x)$ in the gradient case.]
{
\includegraphics[width=0.45\textwidth]{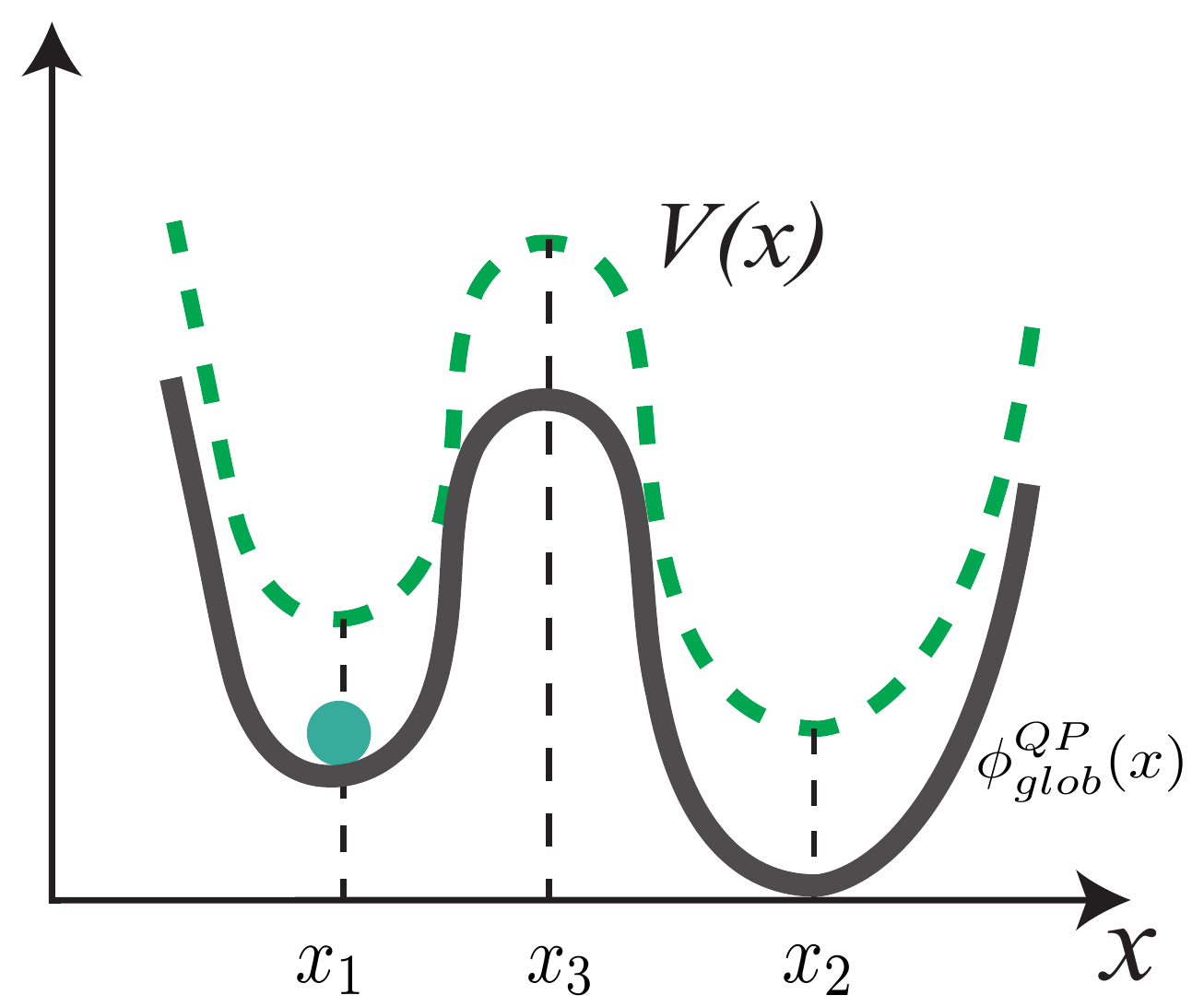}
}
\caption{(Color online). The original potential field and construction of local and global quasi-potential for a gradient system. The subfigure (a) shows the original potential field $V(x)$. The subfigures (b) and (c) show the constructed local quasi-potential starting from metastable states $x_1$ and $x_2$, respectively.  In subfigure (d), the green dashed line is the original potential $V(x)$ and the gray solid line is the global quasi-potential $\phi^{QP}_{glob}(x)$.}
\label{fig:sticking}
\end{figure}

Consider a one dimensional Brownian dynamics with double-well potential $V(x)$
\begin{equation}
dX_{t}=-\nabla V(X_{t})dt+\sqrt{2\varepsilon}dW_{t},
\label{eqn:gradsystem}
\end{equation}
where we assume $V(x)$ has two local minimum points $x_{1}$, $x_{2}$ with $V(x_{1})<V(x_{2})$, and one local maximum point $x_{3}$ in between. A schematics of $V(x)$ is depicted in Fig.~\ref{fig:potential}. In the deterministic version, we have two stable states $x_{1},x_{2}$ and one unstable states $x_{3}$.
 
The local quasi-potential with respect to $x_{1}$ and $x_{2}$ can be obtained by solving the variational problem directly, with details presented in Appendix B. We have 
 \begin{equation*}
\phi^{QP}_{loc}(x;x_{1})=
\begin{cases}
V(x)-V(x_{1}),  & x<x_{3}, \\
V(x_{3})-V(x_{1}), & x_{3}\leq x \leq x_{2}, \\
V(x)+V(x_{3})-V(x_{1})-V(x_{2}) , & x> x_{2},
\end{cases}
\end{equation*}
and similarly
\begin{equation*}
\phi^{QP}_{loc}(x;x_{2})=
\begin{cases}
V(x)+V(x_{3})-V(x_{1})-V(x_{2}), & x<x_{1}, \\
V(x_{3})-V(x_{2}), & x_{1}\leq x \leq x_{3}, \\
V(x)-V(x_{2}), & x>x_{3}.
\end{cases}
\end{equation*}

The general methodology and theoretical results for sticking local quasi-potentials into a global one appear in the book \cite{freidlin2012random}. However in this simple one dimensional example with only two stable points, the strategy is quite straightforward:
\begin{itemize}
\item Step 1. Cut out the parts of the local potential outside of the attraction basin of the starting stable point $x_{1}$ or $x_{2}$.

\item Step 2. Paste the processed local potentials together through the unstable point $x_{3}$.

\item Step 3. Shift the obtained potential such that the minimum of the global quasi-potential is 0.
\end{itemize} 
The Step 3 is not necessary in general since only the difference of potential matters for a dynamical system. Express the above procedure in a mathematical way we have 
\begin{equation}\label{eq:QPGlobExam}
\phi^{QP}_{glob}(x)=\min\Big\{\phi^{QP}_{loc}(x;x_{1})+V_{2,1}, \phi^{QP}_{loc}(x;x_{2})+V_{1,2}\Big\}-\min\{V_{1,2},V_{2,1}\}
\end{equation}
where $V_{1,2}=V(x_{3})-V(x_{1})$ denotes the barrier height from stable point $x_{1}$ to $x_{2}$ and $V_{2,1}=V(x_{3})-V(x_{2})$ denotes the barrier height from stable point $x_{2}$ to $x_{1}$. In the current example, Eq.~\eqref{eq:QPGlobExam} can be further simplified since  we have assumed $V(x_{3})>V(x_{1})>V(x_{2})$, 
\begin{equation*}
\phi^{QP}_{glob}(x)=\min\Big\{\phi^{QP}_{loc}(x;x_{1})+V_{2,1}-V_{1,2}, \phi^{QP}_{loc}(x;x_{2})\Big\}=V(x)-V(x_{2}).
\end{equation*}
Hence the global quasi-potential of system (\ref{eqn:gradsystem}) is just the shift of the real potential $V(x)$  by $-V(x_{2})$. The construction procedure of the global quasi-potential is schematically shown in Fig.~\ref{fig:sticking}. A concrete example of constructing quasi-potential in the case of three attractors and the description of ``$\lambda$-surgery and pasting''  is presented in \cite{ge2012landscapes}.

\subsection{Force Decomposition: HJE and Orthogonality}\label{sec:HJE}

As in Section \eqref{sec:ForceDecNESS}, we will investigate the decomposition of the force $b(x)$ in terms of the global quasi-potential. We will obtain an $\varepsilon$-independent decomposition of $b(x)$, which can be viewed as the limit of Eq. \eqref{eqn:wangdecom}. This decomposition is particularly useful for describing the optimal transition path between meta-stable states.

Substituting the well-known WKB ansatz \cite{Bender1999Book}
$$P_{ss}(x)=\exp\Big(-\frac{\phi(x)}{\varepsilon}+\phi_{0}(x)+\varepsilon\phi_{1}(x)+\cdots\Big)$$
 into the steady Fokker-Planck equation (\ref{eqn:steady state}) and collecting leading order terms, we arrive at a steady Hamilton-Jacobi equation for $\phi(x)$
\begin{equation}\label{eqn:qphje}
[b(x)+D(x)\nabla\phi(x)]\cdot\nabla\phi(x)=0,
\end{equation}
which we simply denote as
\begin{equation}\label{eqn:HJE}
H(x,\nabla \phi)=0.
\end{equation}
Here $H(x,p)$ is exactly the Hamiltonian defined in \eqref{eq:Hamilton}. The relation (\ref{eqn:ssqp}) tells us the fact
$$\phi(x)=\phi^{QP}(x),$$ 
thus $\phi^{QP}(x)$ satisfies the same Hamilton-Jacobi equation \eqref{eqn:qphje}. This point can be also obtained from the Hamilton-Jacobi theory for the variational problem \eqref{eqn:qp} in classical mechanics \cite{landau1960classical}. 

Now we can have the decomposition 
\begin{equation}\label{eqn:fdqp}
b(x)=-D(x)\nabla\phi^{QP}(x)+\ell(x),
\end{equation}
and the orthogonality condition
\begin{equation}
\ell(x)\cdot\nabla\phi^{QP}(x)=0
\end{equation}
holds by \eqref{eqn:qphje}. We note here that the decomposition \eqref{eqn:fdqp} is the zero noise limit of \eqref{eqn:wangdecom} because
$$\varepsilon \phi^{PL}(x)\rightarrow \phi^{QP}(x),\quad \varepsilon \nabla\cdot D(x)\rightarrow 0$$
and $J_{ss}(x)/P_{ss}(x)\rightarrow \ell(x)$ by the decomposition equality of $b(x)$. When the system is at equilibrium, i.e. $J_{ss}(x)=0$, we have $\ell(x)=0$ and thus $b(x)=-D(x)\nabla \phi^{QP}(x)$. In general $\ell(x)$ is not zero and the non-equilibrium effect exists. We comment that the normal decomposition proposed in \cite{zhou2012quasi} is a special case of Eq.  (\ref{eqn:fdqp}) by taking $D(x)=I$.

We take the oscillatory biological dynamics to illustrate the use of force decomposition in the framework of quasi-potential theory. Following the arguments in \cite{ge2012landscapes}, $\phi^{QP}(x)$ is constant along a limit cycle $\Gamma$. This can be shown by noting  
\begin{equation*}
\oint_{\Gamma}\nabla\phi^{QP}\cdot dl=0
\end{equation*} 
and the fact
\begin{equation}\label{eq:QPLimitCycle}
\nabla\phi^{QP}\cdot dl=\frac{\nabla\phi^{QP}\cdot b(x)}{|b(x)|}=-\frac{(\nabla\phi^{QP})^{t}D(x)\nabla\phi^{QP}}{|b(x)|}\leq 0,
\end{equation}
which indicates that $\nabla\phi^{QP}\cdot dl \equiv 0$ and thus $\phi^{QP}(x)$ is constant on $\Gamma$. Furthermore, we have $\nabla \phi^{QP}(x)=0$ on $\Gamma$ by \eqref{eq:QPLimitCycle} if the non-degeneracy condition of $D$ is assumed. Hence from force decomposition (\ref{eqn:fdqp}), we have $b(x)=\ell(x)$ on $\Gamma$, suggesting that the oscillatory biological system is completely driven by the non-gradient force $\ell(x)$ on the limit cycle. However, the potential landscape $\phi^{QL}(x)$ is generally not constant along the limit cycle due to the finite size effect. This phenomenon is explicitly exhibited during the landscape study for budding yeast cell cycle \cite{lv2015energy}. 

The decomposition \eqref{eqn:fdqp} is also useful to characterize the optimal transition path between meta-stable states in the small noise case. Consider two neighboring meta-stable states $x_{0}$ and $x_{1}$ separated by the basin boundary $\Gamma$ with unique saddle $x^{*}$. We aim to find the optimal transition path $\psi(t)$ from $x_{0}$ to $x_{1}$. We have
\begin{align}\label{eq:NonGradFWM}
  S_T[\psi] = & \frac{1}{4}\int_0^T |\dot{\psi}+D(\psi)\nabla \phi^{QP}(\psi) -\ell(\psi)|_{D}^2dt\nonumber \\
  = & \frac{1}{4}\Bigg[\int_0^{T_*} + \int_{T_*}^T  ~|\dot{\psi}+D(\psi)\nabla \phi^{QP}(\psi) - \ell(\psi)|_{D}^2~dt\Bigg]\nonumber \\
  = & \frac{1}{4}\int_0^{T_*} |\dot{\psi}-D(\psi)\nabla \phi^{QP}(\psi) - \ell(\psi)|_{D}^2dt\nonumber\\
  &  +
  \int_0^{T_{*}} \langle \dot{\psi}-\ell(\psi), D(\psi)\nabla \phi^{QP}(\psi)\rangle_{D} dt\nonumber \\
  &+\frac{1}{4}\int_{T_*}^T |\dot{\psi}+D(\psi)\nabla V(\psi) - \ell(\psi)|_{D}^2dt\nonumber \\
 \geq & \int_0^{T^{*}} \dot{\psi}^{t}\nabla \phi^{QP}(\psi)dt = \phi^{QP}(\tilde{x}) - \phi^{QP}(x_{0}),
\end{align}
where $T^{*}$ and $\tilde{x}$ are the time and position that the path cross the basin boundary $\Gamma$, respectively. Here we utilized the notation of weighted inner-product and  norm
$$\langle x,y\rangle_{D} = x^{t}D^{-1}y,\quad |x|^{2}_{D}=\langle x,x\rangle_{D}.$$ 
 The minimization of the Freidlin-Wentzell functional requires that $\tilde{x}= x^{*}$ in  \eqref{eq:NonGradFWM}, the point with lowest quasi-potential on the basin boundary,  and the other two integrals are zero.  This means that the optimal transition path is composed of two segments:
 $$
 \begin{array}{ll}
\text{Uphill path:} & 
\renewcommand{\arraystretch}{1.2}\left\{
\begin{array}{l}
\dot\psi = D(\psi)\nabla \phi^{QP}(\psi)+\ell(\psi),\\
\psi(-\infty)=x_{0},~\psi(\infty)=x^{*},
\end{array}\right.\\
&\\
\text{Downhill path:} &
\renewcommand{\arraystretch}{1.2}\left\{
\begin{array}{l}
 \dot\psi = -D(\psi)\nabla \phi^{QP}(\psi) +\ell(\psi) = b(\psi),\\
\psi(-\infty)=x^{*},~\psi(\infty)=x_{1}.
\end{array}\right.
\end{array}
$$
where we have two bi-infinite boundary value problems because $x_{1},~x_{2}$ and $x^{*}$ are all stationary points of the corresponding dynamics.

Finally we comment that the global quasi-potential can be also numerically computed by solving the Hamilton-Jacobi equation \eqref{eqn:qphje} \cite{cameron2012finding,lu2014construction}. But the curse of dimensionality is always an issue in higher dimensions.

The basic concepts and properties related to quasi-potential theory is summarized in Table \ref{table:qp}. 

\begin{table}[h]
\centering
\rowcolors{1}{dark}{light}
\renewcommand\arraystretch{2}

\begin{tabular}{p{5cm} p{9 cm}}
\multicolumn{2}{c}{\bf Realization of Waddington's Metaphor Candidate II: Quasi-Potential} \\

& $\phi^{QP}_{loc}(x;x_{0})=\inf\limits_{T>0}\inf\limits_{\psi(0)=x_{0},\psi(T)=x}\int_{0}^{T}L^{FW}(\psi,\dot{\psi})dt$ (Transition Path) \\

\rowcolor{light}\multirow{-2}{*}{\bf Definition} & $\phi_{glob}^{QP}(x)=-\lim\limits_{\varepsilon\to 0}\varepsilon\ln P_{ss}(x)$ (Steady State Distribution) \\

\rowcolor{dark}& Geometric Minimum Action Method (gMAM) or Solve Hamilton-Jacobi Equation \\
\rowcolor{dark}\multirow{-2}{*}{\bf Numerical Strategy} & Stick the local quasi-potentials together into the global quasi-potential\\

\rowcolor{light}{\vspace{0.02cm}\bf Force Decomposition} & $b(x)=-D(x)\nabla\phi^{QP}(x)+\ell(x)$ and the orthogonality between gradient and non-gradient term $\quad \langle \ell(x),\nabla\phi^{QP}(x)\rangle=0$, where $\ell(x)$ reflects the NESS nature of the system \\

\rowcolor{dark}{\vspace{0.5cm}\bf Connection with Potential Landscape} & $\lim\limits_{\varepsilon \to 0} \varepsilon\phi^{PL}(x)=\phi^{QP}(x)$. The quasi-potential is a good approximation to the potential landscape when noise is small, where the numerical simulation for $\phi^{PL}(x)$ is inefficient.

$\lim\limits_{\varepsilon \to 0}J_{ss}(x)/P_{ss}(x) =\ell(x)$. The force decomposition based on quasi-potential is the limit version of force decomposition based on potential landscape. 
\end{tabular}

\caption{Summary for the quasi-potential $\phi^{QP}(x)$.}
\label{table:qp}
\end{table}

\section{SDE Decomposition and A-type Integral}

Motivated by the fluctuation-dissipation theorem \cite{zhu2004robustness}, P. Ao and his coworkers performed the SDE decomposition to obtain the underlying potential function $\phi^{AO}(x)$ \cite{ao2004potential} and proposed the so-called A-type integral interpretation of the SDEs \cite{yuan2012beyond}. Not only is the force decomposed in this theoretical framework, some additional restrictions on the decomposed force and stochastic terms are also posed from physical arguments. However, there are seldom mathematical studies on these concepts.  In this section we will show that the potential function $\phi^{AO}(x)$ in the decomposed SDE is nothing but the quasi-potential $\phi^{QP}(x)$ under reasonable conditions. Furthermore, some ambiguities in Ao's SDE decomposition theory such as the existence and uniqueness of the decomposition will also be pointed out and clarified via the connection with the quasi-potential theory. We will show that in general the SDE decomposition is not unique in high dimensions ($n\geq 3$). Therefore the A-type integral is only appropriately defined to a {\it given decomposed form} of SDEs instead of the original one.

\subsection{SDE Decomposition and the Potential}

In \cite{ao2004potential}, it is claimed that the SDEs  (\ref{eqn:original}) can be transformed into an equivalent decomposed form 
\begin{equation}\label{eqn:transform}
[S(X_{t})+A(X_{t})]dX_{t}=-\nabla\phi^{AO}(X_{t})dt+\tilde{\sigma}(X_{t})dW_{t},\quad  \tilde{\sigma}(x)\tilde{\sigma}(x)^{t}=2\varepsilon S(x),
\end{equation}
where $S(x)$ is a positive semi-definite matrix, $A(x)$ is an anti-symmetric matrix and $\phi^{AO}(x)$ is the desired potential function. 

In terms of physical interpretation, the stochastic process of decomposition form (\ref{eqn:transform}) can be related to the following physical process with frictional and Lorentz forces
\begin{equation}\label{eqn:LE}
\left\{
\begin{array}{l}
dX_{t}=V_{t}dt \\
mdV_{t}=-[S(X_{t})+A(X_{t})]V_{t}dt-\nabla\phi^{AO}(X_{t})dt+\tilde{\sigma}(X_{t}) dW_{t},~ \tilde{\sigma}(x)\tilde{\sigma}(x)^{t}=2\varepsilon S(x)
\end{array}\right.
\end{equation}
as the mass of the particle $m$ tends to zero. In high dimensional case ($n>3$), $-S(x)V_{t}$ is the generalization of frictional force, $-A(x)V_{t}$ is the generalization of Lorentz force and $ \tilde{\sigma}(x)\tilde{\sigma}(x)^{t}=2\varepsilon S(x)$ is the generalization of Einstein relation in Langevin dynamics \cite{yuan2012beyond}.

To mathematically execute the transformation from (\ref{eqn:original}) to \eqref{eqn:transform} in practice, some conditions are imposed on $S(x)$ and $A(x)$. By solving these conditions either analytically or numerically, the SDE decomposition as well as potential function $\phi^{AO}(x)$ is thought to be available \cite{ao2004potential}. In this construction, Eqs.  (\ref{eqn:original}) and  (\ref{eqn:transform}) are related by the relationship $[S(x)+A(x)]b(x)=-\nabla\phi^{AO}(x)$ and $[S(x)+A(x)]\sigma(x)=\tilde{\sigma}(x)$. Inserting these expressions into $\nabla\times[-\nabla\phi^{AO}(x)]=0$ and $\tilde{\sigma}(x)\tilde{\sigma}(x)^{t}=2\varepsilon S(x)$ will yield the following constraints on $S(x)$ and $A(x)$:
\begin{subequations}\label{eqn:condition}
\begin{gather}
\nabla\times[(S(x)+A(x))b(x)]=0, \label{subeq:asym} \\
[S(x)+A(x)]D(x)[S(x)-A(x)]=S(x),\label{subeq:sym}
\end{gather}
\end{subequations}
in which $\nabla\times f$ is defined as the $n\times n$ anti-symmetric matrix $(\nabla\times f)_{ij}=\partial_{i}f_{j}-\partial_{j}f_{i}$ for $f\in\mathbb{R}^{n}$. Hence Eqs.~(\ref{subeq:asym}) and (\ref{subeq:sym}) form a nonlinear PDE system with  $(n^2-n)/2$ and $(n^2+n)/2$ equations, respectively.

It is claimed in \cite{ao2004potential} that for given $b(x)$ and $D(x)$, the above $n^{2}$ conditions can determine $n^{2}$ unknown functions in $S(x)$ which is symmetric with $(n^2+n)/2$ unknowns, and $A(x)$ which is anti-symmetric with $(n^2-n)/2$ unknowns. Having solved $S(x)$ and $A(x)$, the potential function $\phi^{AO}(x)$ in (\ref{eqn:transform}) is then given by
\begin{equation}\label{eq:DefPhiAo}
\nabla\phi^{AO}(x)=-[S(x)+A(x)]b(x)
\end{equation} 
as the consequence. However, this assertion needs further mathematical justification since solving the nonlinear PDE system (\ref{subeq:asym})-(\ref{subeq:sym}) is not so straightforward.

\subsection{Steady State Distribution: A-type Integral Interpretation}

One of the key parts of the proposal \cite{ao2004potential} is that the stochastic integral in the decomposed SDE (\ref{eqn:transform}) should be interpreted as the so-called A-type integral beyond Ito or Stratonovich framework \cite{yuan2012beyond}, which is defined as follows.
 
Assume that $[S(x)+A(x)]$ is invertible in Eq.  (\ref{eqn:transform})  and denote $G(x)=[S(x)+A(x)]^{-1}$. From (\ref{subeq:sym}) we have $G(x)+G^{t}(x)=2D(x)$, or $G(x)=D(x)+Q(x)$ where $Q(x)$ is an anti-symmetric matrix. The  A-type Fokker-Planck equation  for the decomposed process\footnote{It is also claimed in \cite{yuan2012beyond} that the A-type integral interpretation can be equivalently applied to the original process (\ref{eqn:original}). However, the results in Section \ref{sec:nonunique}  suggest that the A-type Fokker-Planck equation  for (\ref{eqn:original}) might not be well-determined if the dimension $n\geq 3$.} (\ref{eqn:transform}) can be derived from zero-mass limit of the extended system (\ref{eqn:LE}) \cite{yin2006existence}: 
\begin{equation}
\partial_{t}\rho=\nabla\cdot G(\varepsilon\nabla+\nabla\phi^{AO})\rho=-\nabla\cdot(b\rho)+\varepsilon\nabla\cdot(D+Q)\nabla\rho.
\label{eqn:aint}
\end{equation}
In one dimensional case ($n=1$), this Fokker-Planck equation corresponds to the right-most endpoint stochastic integral interpretation of Eq.  (\ref{eqn:original}), but there is no explicit stochastic integral interpretation of it in higher dimensions.  

One feature about the Fokker-Planck equation of A-type Integral is that its steady state distribution is of Boltzman form:
\begin{equation*}
P_{ss}(x)=\frac{1}{Z_{\varepsilon}}\exp\Big({-\frac{\phi^{AO}(x)}{\varepsilon}}\Big)
\end{equation*}
with the potential function $\phi^{AO}(x)$ appearing in the decomposition. Hence in this case we obtain
\begin{equation}\label{eq:phiao}
\phi^{AO}(x)=-\varepsilon\ln P^{\text{A-type}}_{ss}(x) -\varepsilon\ln Z_{\varepsilon}.
\end{equation}
The first looking on \eqref{eq:phiao} is reminiscent of potential landscape $\phi^{PL}(x)$. But a careful comparison tells us that the steady state distribution $P^{\text{A-type}}_{ss}(x)$ is totally different from the $P_{ss}(x)$ in \eqref{eqn:pl} because of different interpretations of SDEs. This often brings confusions in the literature. Furthermore, we will show that $\phi^{AO}(x)$ is nothing but the quasi-potential $\phi^{QP}(x)$.

\subsection{Force Decomposition Revisited}

By revisiting the SDE decomposition theory from force decomposition perspective, we will show that $\phi^{AO}(x)$ coincides with $\phi^{QP}(x)$ under reasonable conditions. This implies that for a given SDE (\ref{eqn:original}): (i) Its corresponding $\phi^{AO}(x)$ can be interpreted as a quasi-potential landscape; (ii) Its quasi-potential $\phi^{QP}(x)$ corresponds to the gradient term in a certain decomposed SDE with respect to SDE (\ref{eqn:original}).  As a corollary, conclusions on the existence and uniqueness issues of the SDE decomposition theory can be drawn.

\subsubsection{Connection with Quasi-Potential }\label{sec:nonunique}

Recall that if we denote $G(x)=[S(x)+A(x)]^{-1}$, then the relation (\ref{subeq:sym}) yields $G(x)=D(x)+Q(x)$ where $Q(x)$ is an anti-symmetric matrix.  Now we can decompose the force $b(x)$ with the form
\begin{gather*}
\begin{split}
b(x)&=-G(x)\nabla\phi^{AO}(x) \\
&=-[D(x)+Q(x)]\nabla\phi^{AO}(x) \\
&= -D(x)\nabla\phi^{AO}(x)+\ell(x)
\end{split}
\end{gather*}
where $\ell(x)=-Q(x)\nabla\phi^{AO}(x)$. Since $Q(x)$ is anti-symmetric, we have
\begin{equation*}
\ell(x)\cdot\nabla\phi^{AO}(x)=-(\nabla\phi^{AO})^{t}Q\nabla\phi^{AO}=0
\end{equation*}
Therefore just as the quasi-potential $\phi^{QP}(x)$,  the potential function $\phi^{AO}(x)$ also satisfies the Hamilton-Jacobi equation 
\begin{equation}\label{eqn:hje}
[b(x)+D(x)\nabla\phi^{AO}(x)]\cdot\nabla\phi^{AO}(x)=0
\end{equation}

The fact that $\phi^{AO}(x)$ and $\phi^{QP}(x)$ share the same partial differential equation (\ref{eqn:hje}) tells us that they are indeed the same function, at least when $b(x)$ has only one stationary stable state and no other attractors since there are multiple solutions of the HJE in general.
Meanwhile, we do not know how to regularize to select the reasonable solution of \eqref{eqn:hje} based on the original definition of $\phi^{AO}(x)$. But we will accept the choice that $\phi^{AO}(x)$ is the noise vanishing limit of \eqref{eqn:hje} as the quasi-potential $\phi^{QP}(x)$ in this paper.

 The result \eqref{eqn:hje} also implies that the construction of $\phi^{AO}(x)$ can be achieved by the same strategy as discussed for the quasi-potential $\phi^{QP}(x)$, while the naive method by utilizing the definition \eqref{eq:DefPhiAo} directly is not a feasible approach because the solution of $S(x)$ and $A(x)$ may be an even harder problem.  On the contrary, we will instead study the decomposition and determine the corresponding $S^{QP}(x)$ and $A^{QP}(x)$ through the obtained quasi-potential $\phi^{QP}(x)$, i.e. $\phi^{AO}(x)$. With this perspective, we define
\begin{equation}\label{eqn:recons}
[S^{QP}(X_{t})+A^{QP}(X_{t})]dX_{t}=-\nabla\phi^{QP}(X_{t})dt+\tilde{\sigma}(X_{t})dW_{t}, \quad \tilde{\sigma}(x)\tilde{\sigma}(x)^{t}=2\varepsilon S^{QP}(x),
\end{equation}
which we called the {\it reconstruction} of SDE decomposition starting from quasi-potential. Through this reconstruction, the quasi-potential  $\phi^{QP}(x)$ can be reinterpreted as $\phi^{AO}(x)$, which also indicates that $\exp(-\phi^{QP}(x)/\varepsilon)$ can serve as the steady-state distribution under the A-type stochastic integral interpretation of Eq.  (\ref{eqn:recons}).  

Our theoretical results on the reconstruction deal with arbitrary solution $\phi(x)$ of Hamilton-Jacobi equation, which is not limited to the quasi-potential $\phi^{QP}(x)$. The existence of the reconstruction (\ref{eqn:recons}) starting from $\phi(x)$ is guaranteed by the following theorem:
\begin{theorem}
Suppose $D(x)$ is not singular and $\phi(x)$ is the solution of Hamilton-Jacobi equation  (\ref{eqn:hje}). If $b(x)$ and $\nabla\phi(x)$ have the same zeros, then there exist a positive definite matrix $S(x)$ and an anti-symmetric matrix $A(x)$ such that $[S(x)+A(x)]b(x)=-\nabla\phi(x)$ and $[S(x)+A(x)]D(x)[S(x)-A(x)]=S(x)$.
\label{thm:main}
\end{theorem}

We also discovered that in general the constructed $S(x)$ and $A(x)$ are not unique in high dimensions ($n\geq3$). Moreover, this under-determination of $S$ and $A$ can be also quantitatively characterized:
\begin{theorem}
Suppose $D(x)$ is not singular and $\phi(x)$ is the solution of Hamilton-Jacobi equation (\ref{eqn:hje}). If $\phi(x)$ is also nonsingular (i.e. $\nabla\phi(x)\not=0 $) for fixed $x\in\mathbb{R}^{n}$, then $S(x)$ and $A(x)$ in Theorem \ref{thm:main} have the degrees of freedom $(n-1)(n-2)/2$.
\label{thm:sub}
\end{theorem}

The detailed proof is presented in the Appendix C. Here we demonstrate the main idea in the reconstruction procedure appeared in the proof, which involves three steps:

\begin{itemize}
\item Step~1. Finding a solution $\phi(x)$ of Hamilton-Jacobi equation (\ref{eqn:hje}) with appropriate boundary conditions. 

\item Step~2. Constructing a matrix function $G(x)$ such that $G(x)\nabla\phi(x)=-b(x)$ and $G(x)+G^{T}(x)=2D(x)$. We can show that the desired $G(x)$ can be constructed by solving certain linear systems, whose solvability is guaranteed by the conditions in the theorem. The degrees of freedom of $G(x)$ can be also obtained.

\item Step~3. Setting  $S(x)=[G^{-1}(x)+G^{-T}(x)]/2$ and $A(x)=[G^{-1}(x)-G^{-T}(x)]/2$, thus obtaining the decomposed form in Eq.  (\ref{eqn:transform}). The invertibility of $G(x)$ is implied by the non-singularity of $D(x)$.
\end{itemize}

Let us remark on the two conditions imposed in Theorem \ref{thm:main}: the non-singularity of diffusion matrix $D(x)$ and the common-zero assumption of $b(x)$ and $\nabla\phi(x)$.  It can be found in the proof that the non-singularity of $D(x)$ is just a technical condition to ensure the invertibility of the constructed $G(x)$ in the second step. In practice, as long as the solved $G(x)$ is invertible, this assumption on $D(x)$ can be removed.  For the common-zero assumption, in the first place one can show that if $b(x_{0})=0$, then $\nabla\phi(x_{0})=0$ provided that $\det D(x_{0})\neq 0$.  The violation of common-zero assumption mostly happens in the case $b(x_{0})\neq 0$ and $\nabla\phi(x_{0})=0$. From $[S(x_{0})+A(x_{0})]b(x_{0})=-\nabla\phi(x_{0})$ we know $S(x_{0})+A(x_{0})$ is degenerate, implying that the A-type Integral Fokker-Planck equation  is not well-defined at $x_{0}$. A related example will be demonstrated in Section \ref{sec:example}.

\subsubsection{Existence and Uniqueness Issue of SDE Decomposition}

Our theoretical results on the SDE decomposition lead to the discussion about rigorous mathematical aspects of Ao's proposal. 

One fundamental theoretical issue is the existence and uniqueness of the SDE decomposition. Starting from SDE (\ref{eqn:original}), the quasi-potential function $\phi^{QP}(x)$ can be constructed. As long as $\phi^{QP}(x)$ satisfies the common-zero assumption (which can be viewed as an inherent property of the SDE), the existence of the decomposition can be established as the corollary of Theorem \ref{thm:main}. However, Theorem \ref{thm:sub} suggests that when $n\geq 3$, the original SDE might be decomposed into a family of different SDEs in form (\ref{eqn:transform}) satisfying the restrictions in (\ref{eqn:condition}) (these SDEs share the same potential function $\phi^{QP}(x)$), indicating that the imposed conditions in the theory do not uniquely determine the decomposition. 

These results clarify that it is more appropriate to apply the A-type stochastic integral interpretation to the decomposed SDEs (\ref{eqn:transform}) rather than the original SDE (\ref{eqn:original}), since there might be a family of different $G(x)$ corresponding to the same original SDEs, which renders the Fokker-Planck equation  undetermined (c.f. the example in Section \ref{sec:aoexample}).

Therefore we conclude that the potential function $\phi^{AO}(x)$ and the decomposition matrix $S(x),A(x)$ and $G(x)$ should be analyzed separately in the SDE decomposition theory. In the A-type integral framework,  $\phi^{AO}(x)$ (which is shown to be consistent with $\phi^{QP}$ in many situations) determines the steady state distribution, while $S(x),A(x)$ and $G(x)$ (which are not uniquely determined in general) reveals the relaxation behavior of probability evolution in Fokker-Planck equation. It is  interesting to note that for a given SDE (\ref{eqn:original}), there may exist various relaxation processes leading to the same invariant distribution under A-type integral interpretation, which suggests that the potential function $\phi(x)$ serves as a more characteristic quantity for SDE (\ref{eqn:original}) rather than $S(x)$ and $A(x)$.

\begin{table}[p]
\centering
\rowcolors{1}{dark}{light}
\renewcommand\arraystretch{2}

\begin{tabular}{p{4.9 cm} p{9.1 cm}}
\multicolumn{2}{c}{\bf Realization of Waddington's Metaphor Candidate III: SDE Decomposition} \\

& Decompose the original SDE into $[S(X_{t})+A(X_{t})]dX_{t}=-\nabla\phi^{AO}(X_{t})+\tilde{\sigma}(X_{t})dW_{t}$, with the restriction $\tilde{\sigma}(x)\tilde{\sigma}(x)^{t}=2\varepsilon S(x)$  \\
\rowcolor{light}\multirow{-2}{*}{\bf Definition} & The potential function is defined by $\nabla\phi^{AO}(x)=-[S(x)+A(x)]b(x)$ (Force Decomposition) \\

\rowcolor{dark}{\bf Numerical Strategy}& Solve $n^2$ non-linear PDEs (not practical in general) \\

\rowcolor{light}{\bf Steady State Distribution} & Interpreted under  A-type integral framework, $\nabla\phi^{AO}(x)=-\varepsilon\nabla\ln P^{\text{A-type}}_{ss}(x)$ \\

\rowcolor{dark}{\vspace{0.02cm}\bf Force Decomposition} & $b(x)=-D(x)\nabla\phi^{AO}(x)+\ell(x)$ plus the orthogonality between gradient and non-gradient term $\quad \langle \ell(x),\nabla\phi^{AO}(x)\rangle=0$, where $\ell(x)=Q(x)\phi^{AO}(x)$ and $Q(x)$ is an anti-symmetric matrix \\

\rowcolor{light}{\vspace{0.005cm}\bf Connection with Quasi-Potential} & $\phi^{AO}(x)$ coincides with $\phi^{QP}(x)$ in broad situations. Although the two functions are interpreted under different frameworks, as the landscape function they are the same.\\

\rowcolor{dark}{\bf Connection with Potential Landscape} & $\lim\limits_{\varepsilon \to 0} \varepsilon\nabla\phi^{PL}(x)=\nabla\phi^{AO}(x)$. \\

\rowcolor{light}{\vspace{0.01cm}\bf Existence \& Uniqueness of Decomposition} & The existence of Ao's SDE decomposition for general diffusion process can be guaranteed under the reasonable conditions stated in Theorem 1 while in high dimensional case $(n\geq 3)$ the decomposition $(S,A,Q)$  is not unique in general.
\end{tabular}
\caption{Summary for the potential function $\phi^{AO}(x)$ constructed in SDE decomposition theory. }
\label{table:ao}
\end{table}

We summarize our explorations and findings on the SDE decomposition theory in Table \ref{table:ao}.

\subsection{Illustrative Examples}\label{sec:aoexample}

In this subsection, we will provide two examples to concretely illustrate our theoretical results on the SDE decomposition. 

\subsubsection{Two-dimensional Case: Uniqueness of Decomposition}

Let us consider the 2-D diffusion process with constant noise amplitude:
\begin{equation*}
\begin{cases}
dX_t=b_{1}(X_{t},Y_{t})+\sqrt{2\varepsilon}dW^{1}_{t}, \\
dY_t=b_{2}(X_{t},Y_{t})+\sqrt{2\varepsilon}dW^{2}_{t}, \\
\end{cases}
\end{equation*}
where $W^{j}_{t}~(j=1,2)$ are independent Brownian motions. Suppose that $\phi(x,y)$ is the solution of the HJE and the non-singularity and common-zero assumptions both hold, we know from Theorem \ref{thm:sub} the degrees of freedom is zero, hence the decomposition is unique at non-singular points. While at the singular point $G$ could be anything by reconstruction procedure, thus we just define it as the limit  of the values of $G$ at nonsingular points.  To gain more intuitions, let us show this  through direct verification.

From $G+G^{T}=2I$ we know the mapping $G(x,y)$ must take the form
\begin{equation*}
G(x,y)=\left(
\begin{matrix}
1 & g(x,y) \\
-g(x,y) &  1 \\
\end{matrix}
\right),
\end{equation*}
and $g(x,y)$ satisfies
\begin{equation}
\begin{cases}
\partial_{x}\phi(x,y)+g(x,y)\partial_{y}\phi(x,y)=-b_{1}(x,y), \\
-g(x,y)\partial_{x}\phi(x,y)+\partial_{y}\phi(x,y)=-b_{2}(x,y). \\
\end{cases}
\label{example2d:g}
\end{equation}
Since the quasi-potential $\phi$ satisfies HJE  (\ref{eqn:hje}), we conclude that (\ref{example2d:g}) has the unique solution
\begin{equation}
g(x,y)=-\frac{b_{1}+\partial_{x}\phi}{\partial_{y}\phi}=\frac{b_{2}+\partial_{y}\phi}{\partial_{x}\phi}.
\end{equation} 
Consequently $S(x)$ and $A(x)$ can be uniquely determined. For concrete examples, one can easily verify that $\phi(x,y)=(x^{2}+y^{2})/2$ and $g(x,y)=-y$ if we specially take $b_{1}(x,y)=-x+y^{2}$ and $b_{2}(x,y)=-y-xy$.

This example also demonstrates the fact that in the SDE decomposition, even if the diffusion matrix $D$ is constant, the decomposed matrix $G$ (or $Q$) can be variable-dependent.

\subsubsection{Three-dimensional Case: Non-uniqueness of Decomposition}

To illustrate the under-determination of $S(x)$ and $A(x)$ from condition (\ref{eqn:condition}) in higher dimensions, let us consider the following SDEs
\begin{equation}
\begin{cases}
dX_t=(-X_t+Y_{t}^{2})dt+\sqrt{2\varepsilon}dW^{1}_{t}, \\
dY_t=(-Y_t-X_{t}Y_{t})dt+\sqrt{2\varepsilon}dW^{2}_{t}, \\
dZ_t=-Z_tdt+\sqrt{2\varepsilon}dW^{3}_{t},
\end{cases}
\label{example3d:sde}
\end{equation}
where $W^{j}_{t}~(j=1,2,3)$ are independent Brownian motions. The quasi-potential is readily solved from HJE with $\nabla\phi^{QP}=(x,y,z)^{T}$.

From the reconstruction procedure and the remarks in Appendix \ref{sec:remark},  we can obtain
\begin{equation*}
G_{\lambda}(x,y,z)=\left(
\begin{matrix}
 1 & -y+\lambda z & -\lambda y  \\
y-\lambda z & 1 & \lambda x \\ 
\lambda y & -\lambda x & 1  \\
\end{matrix}
\right),
\end{equation*}
where $\lambda(x,y,z)$ is arbitrary smooth function of $x,y$ and $z$. Using the relation $S_{\lambda}=(G_{\lambda}^{-1}+G_{\lambda}^{-T})/2$ and $A_{\lambda}=(G_{\lambda}^{-1}-G_{\lambda}^{-T})/2$ we have
\begin{gather*}
S_{\lambda}(x,y,z)=\frac{1}{K(x,y,z)}\left(
\begin{matrix}
 1+\lambda^2 x^2 & \lambda^2 xy & -\lambda xy+\lambda^2 xz  \\
\lambda^2 xy & 1+\lambda^2 y^2 & -\lambda y^2+\lambda^2 yz \\
-\lambda xy+\lambda^2 xz & -\lambda y^2+\lambda^2 yz & 1+(y-\lambda z)^2  \\
\end{matrix}
\right),\\
A_{\lambda}(x,y,z)=\frac{1}{K(x,y,z)}\left(
\begin{matrix}
 0 & y-\lambda z & \lambda y  \\
-y+\lambda z & 0 & -\lambda x \\
-\lambda y & \lambda x & 0  \\
\end{matrix}
\right),
\end{gather*}
where $K(x,y,z)=\lambda^2(x^2+y^2)+(y-\lambda z)^2+1$. Hence the original system is transformed into a family of stochastic processes with decomposition form
\begin{equation}
\Big[S_{\lambda}(X_{t},Y_{t},Z_{t})+A_{\lambda}(X_{t},Y_{t},Z_{t})\Big]d
\left(
\begin{matrix}
X_{t} \\
Y_{t} \\
Z_{t} \\
\end{matrix}
\right)=-\left(
\begin{matrix}
X_{t} \\
Y_{t} \\
Z_{t} \\
\end{matrix}
\right)+\sqrt{2\varepsilon}\sigma_{\lambda}(X_{t},Y_{t},Z_{t})d\left(
\begin{matrix}
W^{1}_{t} \\
W^{2}_{t} \\
W^{3}_{t} \\
\end{matrix}
\right),
\label{example3d:trans}
\end{equation}
where $\sigma_{\lambda}(X_{t},Y_{t},Z_{t})=S_{\lambda}(X_{t},Y_{t},Z_{t})+A_{\lambda}(X_{t},Y_{t},Z_{t})$. It is easy to verify that (\ref{example3d:trans}) satisfies all of the conditions in the SDE decomposition proposal.

Therefore we note that the SDE decomposition for (\ref{example3d:sde}) is not unique because of the arbitrariness of $\lambda(x,y,z)$, which means that given a stochastic differential equation in form (\ref{eqn:original}), there may exist a family of corresponding processes of form (\ref{eqn:transform}). In this example, we know from Eq. (\ref{eqn:aint}) that the corresponding A-type integral Fokker-Planck equation for PDF $\rho_{\lambda}(x,y,z)$  takes the form 
\begin{equation*}
\partial_{t}\rho_{\lambda}=\partial_{x}\rho_{\lambda}-\nabla\cdot(b\rho_{\lambda})-(\nabla\lambda\times\nabla\phi)\cdot\nabla\rho_{\lambda}+\varepsilon\Delta\rho
_{\lambda},
\end{equation*}
where $b=(-x+y^2,-y-xy,-z)^{T}$ and $\nabla\phi=(x,y,z)^{T}$. Hence starting from the same SDE, with different choice of $\lambda(x,y,z)$, we reach different stochastic processes  (\ref{example3d:trans}) and different A-type integral Fokker-Planck equations.

\section{Comparative Study Through a Toy Example}\label{sec:example}

In this section, a simple yet illuminating example will be provided to help us gain better understanding of the landscape construction proposals discussed in previous sections. We consider the following diffusion process defined on the circle $\mathbb{S}[0,1]$:
\begin{equation}
dX_{t}=dt+\sqrt{2\varepsilon}d \tilde{W}_{t},\quad X_{t}\in \mathbb{S}[0,1]
\label{eqn:circle}
\end{equation}
where $\tilde{W}_{t}$ is the Brownian motion on the circle. Physically, it describes a a particle doing uniform circular motion under random perturbations.

In potential landscape theory, let $p(x,t)$ denote the probability density of particle appearing at point $x\in [0,1]$ on the circle  at time $t$. The Fokker-Planck equation  is
\begin{equation*}
\partial_{t} p + \partial_{x}(p-\varepsilon\partial_{x} p)=0, \quad p(x,t)=p(x+1,t).
\end{equation*}
We can obtain the steady state distribution $P_{ss}(x)=1, x\in [0,1]$ and steady state flux $J_{ss}(x)=1$. Therefore the potential landscape satisfies $\phi^{QP}(x)=\ln P_{ss}(x)=0$ on the circle. From force decomposition perspective, the particle is completely driven by the curl term $J_{ss}/P_{ss}=1$, reflecting the non-equilibrium nature of the system. In fact, the uni-direction feature of this system has close relation with the concept of entropy production in non-equilibrium statistics \cite{jiang2004mathematical}. The rotation number of this system is 1, and the entropy production rate is $\varepsilon^{-1}$.

\begin{figure}[htbp]
\centering 
\includegraphics[scale=0.6]{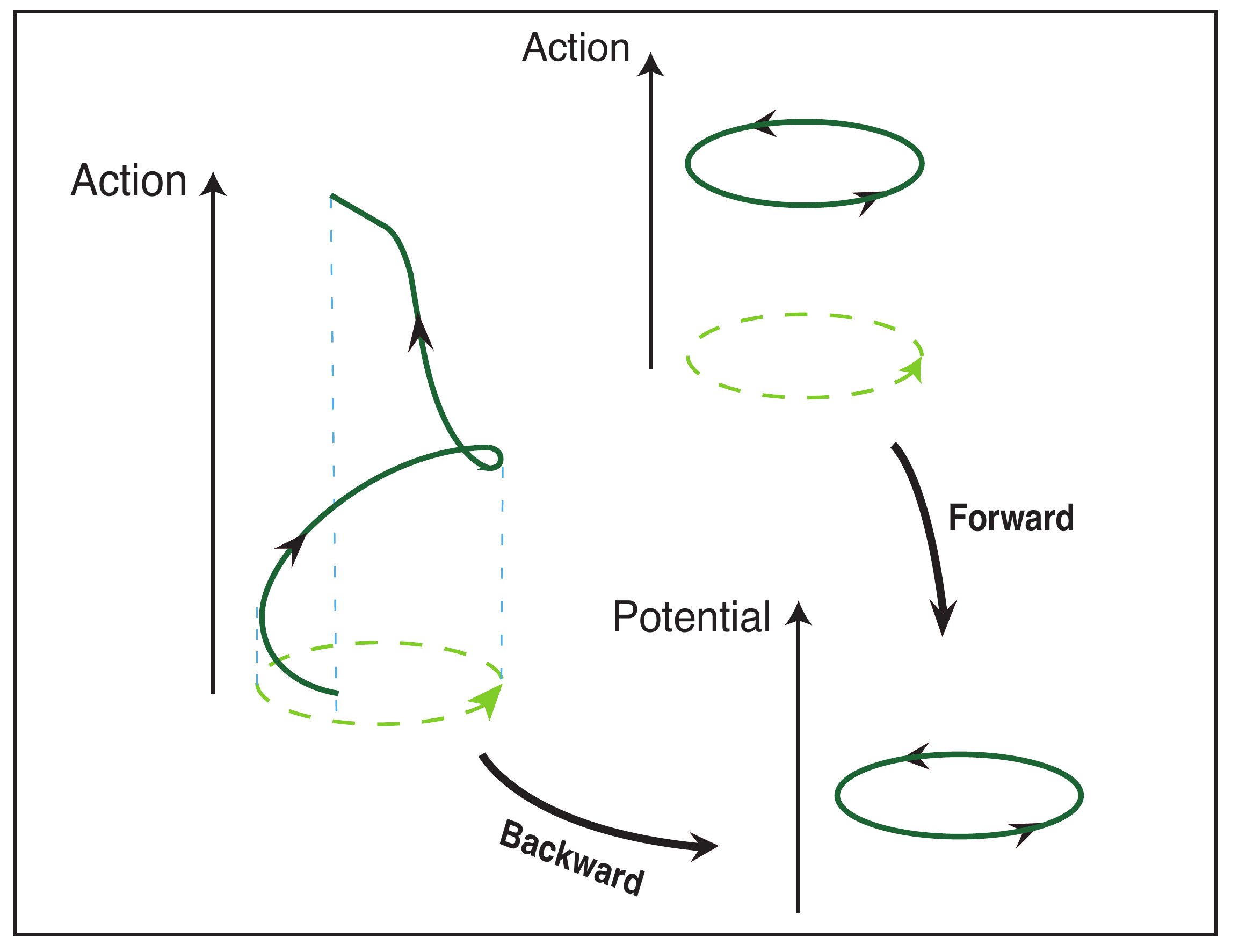}
\caption{(Color online). Constructing the quasi-potential for the diffusion process on the circle modeled by Eq.  (\ref{eqn:circle}). If the path is in the opposite direction of the force term, the action will increase along it. The least action path satisfies the deterministic counterpart of the stochastic process where the action as well as the quasi-potential remains to be zero. This reflects the time irreversibility of the system.}
\label{fig:reverse}
\end{figure}

In the framework of quasi-potential, the landscape can be either computed from Hamilton-Jacobi equation or minimum action approach. The Hamilton-Jacobi equation  is 
\begin{equation*}
(1+\phi'(x))\phi'(x)=0
\end{equation*}
with boundary condition $\phi(x)=\phi(x+1)$, yielding the solution $\phi^{QP}(x)=\text{Constant}$. From the minimum action approach, the least action path $\psi(t)$ connecting any points $x_{1}$ and $x_{2}$ on the circle satisfies the deterministic counterpart $\dot{\psi}(t)=1$ and the action on the path is zero, also indicating that the quasi-potential on the circle should be constant. Interestingly, if we choose the transition path in the opposite direction, then action will increase along the path, which also reflects the system's non-equilibrium property (the time irreversibility). The driving force on the particle is solely the non-gradient term $\ell(x)$ along the circle. This phenomenon is depicted in Fig.~\ref{fig:reverse}. 

Because the quasi-potential $\phi'(x)=0$ on the circle while $b(x)=1$, we cannot apply our results on the connection between the quasi-potential and $\phi^{AO}$ directly as stated in Theorem 1. We will construct the SDE decomposition directly from definition. Assume the SDE decomposition has the form
\begin{equation*}
[s(X_{t})+a(X_{t})]dX_{t}=-\phi'(X_{t})+\tilde{\sigma}(X_{t})d\tilde{W}_{t},\quad X_{t}\in \mathbb{S}[0,1], \quad\tilde{\sigma}^{2}(x)=2\varepsilon s(x).
\end{equation*}
Since $n=1$, we have $a(x)=0$ and $\phi'(x)=s(x)$ in (\ref{eqn:condition}). Moreover, condition (\ref{eqn:condition}) yields $s^2(x)=s(x)$, implying that $s(x)=0$ or $s(x)=1$ by the smoothness of $s(x)$. On the other hand, the boundary condition $\phi(x+1)=\phi(x)$ requires that $\phi'(x)=s(x)=0$. With these facts, we know that the decomposed equation (\ref{eqn:transform}) is not well defined because both sides are zero. Moreover, the A-type integral Fokker-Planck equation  (\ref{eqn:aint}) does not apply in this case. 

The comparison of different proposals to realize Waddington's metaphor for the system modeled by Eq. (\ref{eqn:circle}) is by presented in Table \ref{table:compare}.
\begin{table}[h]
\centering
\rowcolors{1}{dark}{light}
\renewcommand\arraystretch{2}
\begin{tabular}{lp{3cm}p{6.5cm}}
\bf Proposals & \bf Landscape Function & \bf Special Features\\
\bf Potential Landscape & $\phi^{PL}(x)=0$ & The system (rotation number 1, entropy production rate $\varepsilon^{-1}$) is completely driven by non-gradient force. \\
\bf Quasi-Potential & $\phi^{QP}(x)=0$ & The action remains constant along the clockwise path while increases along the anti-clockwise path. \\
\bf SDE Decomposition & $\phi^{AO}(x)=?$ & The both sides of the decomposed SDE are 0 and the corresponding A-type integral Fokker-Planck Equation is ill-defined.
\end{tabular}
\caption{Different realizations of Waddington's metaphor for the simple diffusion process on the circle.}
\label{table:compare}
\end{table}

Although the considered example is just a toy model, it has already been discovered in the study of a cell cycle model \cite{lv2015energy} that similar phenomenon can happen in biological networks where the gradient of quasi-potential vanishes on a manifold. This example informs us that in some non-equilibrium systems, the landscape itself cannot describe the whole picture, thus must be combined with other tools or perspectives such as force decomposition and transition path, to obtain a more comprehensive understanding of the system.

\section{Discussions and Conclusion}

In this paper, we have adopted three perspectives (steady state distribution, transition path and force decomposition) to investigate three existing landscape theories, which can be viewed as quantitative realizations of Waddington's metaphor. The connections between these theories are revealed and some insights are brought out as the consequence of such connections. 

To briefly summarize, we conclude that the quasi-potential $\phi^{QP}(x)$ is the limit of potential landscape $\phi^{PL}(x)$ as the noise strength goes to zero, and the potential $\phi^{AO}(x)$ in SDE decomposition theory coincides with $\phi^{QP}(x)$ in many situations.  We also discover that the condition \eqref{eqn:condition} does not uniquely determine the decomposition. To avoid ambiguity, it will be more reasonable to define the A-type integral  to the decomposed form (\ref{eqn:transform}) rather than the original form.  Next we provide some discussions based on the results in this paper.

\subsection{Numerical Computations}

From numerical perspective, the Monte Carlo simulation is more efficient than the deterministic methods in high dimensions to obtain potential landscape $\phi^{PL}(x)$, while when the noise amplitude is small,  even Monte Carlo simulation becomes costly due to the metastability issue \cite{Frenkel2002Book}. In such circumstance, it is advisable to approximate potential landscape $\phi^{PL}$ by quasi-potential $\phi^{QP}$, which can be computed by gMAM or other numerical methods. The path integral formulation adopted by us to establish the relation between potential landscape and path action in Eq.  (\ref{eqn:pltp}) might imply other MCMC algorithms to compute the potential landscape, especially when the noise amplitude $\varepsilon$ is in the intermediate regime,  where it is too large to adopt quasi-potential approximation while too small to effectively conduct Monte-Carlo simulation. In such cases,  the idea of importance sampling might help to get the estimation of $\phi^{PL}(x)$ effectively.

\subsection{Exchange of Limit Order and Keizer's Paradox}\label{sec:DiscussExch}

Our discussion on the different limit orders appeared in  (\ref{eqn:globallim}) and (\ref{eqn:locallim}) is reminiscent of Keizer's paradox \cite{vellela2007quasistationary}, which involves the inconsistent behavior of deterministic models and chemical master equation models, while our analysis mainly focuses on the deterministic models and diffusion process models.  The limit order in (\ref{eqn:locallim}) corresponds to the behavior of deterministic process ($\varepsilon\to 0$ for fixed $t$) that evolves for a long time ($t\to+\infty$), while the limit order in (\ref{eqn:globallim}) corresponds to the long time behavior of the stochastic process ($t\to+\infty$ for fixed $\varepsilon$) with vanishing noise strength ($\varepsilon\to 0$). When taking the limit, $P^{\varepsilon}(x,t | x_{0},0)$ in the former regime converges to the delta function $\delta(x_{0})$, where $x_{0}$ is the unique local minimum point of local quasi-potential with respect to stable state $x_{0}$. On the contrary, since the global quasi-potential has many local minimum points, $P^{\varepsilon}(x,t | x_{0},0)$ in the latter regime will eventually converge regardless of the initial position $x_{0}$, to the delta function $\delta(x_{m})$,  where $x_{m}$ is the global minimum point of the global quasi-potential. These results are based on the basic intuition that the deterministic system will stay at one specific stable point dependent on the initial position as time goes to infinity, while the long time evolution of the stochastic system modeled by diffusion process will be mainly constrained to the ``most stable state'' as studied in the simulated annealing algorithm. Moreover, the local and global quasi-potential is playing the role as  ``convergence rate'' to the delta functions respectively.

\subsection{Existence and Non-uniqueness of SDE Decomposition} 

Previous results on existence and uniqueness issues about the SDE decomposition was mainly concerned with the Fokker-Planck equations.  The constructive proof provided in \cite{ma2015complete} suggests  the existence of a process in decomposed form whose A-type Fokker-Planck equation coincides with the Ito-type Fokker-Planck equation  for SDE (\ref{eqn:original}). Moreover, Xing's work \cite{xing2010mapping} indicates that these processes are not unique since there is a class of processes which correspond to the same A-type Fokker-Planck equation.

In comparison with previous attempts, our results focus on the existence and uniqueness on the decomposition from (\ref{eqn:original}) to (\ref{eqn:transform}). The existence result for Eq.  (\ref{eqn:transform}) in our paper deals with different processes from the one stated in \cite{ma2015complete} because $Z_{\varepsilon}^{-1}\exp(-\phi^{AO}(x)/\varepsilon)$ is not the steady state distribution of SDE (\ref{eqn:original}) under Ito-type stochastic integral interpretation. Meanwhile, our results also provide a negative answer to the open problem raised in \cite{xing2010mapping}: whether conditions (\ref{eqn:condition}) are sufficient to determine the decomposition uniquely? In cases when $n\geq 3$, we find that there is a class of processes of form (\ref{eqn:transform}) which correspond to the same SDE (\ref{eqn:original}) and their  A-type Fokker-Planck equations are different.
 
The non-uniqueness of SDE decomposition also appears in the construction of Lyapunov functions for dynamical systems \cite{yuan2010constructive}. However, the non-uniqueness there arises from the arbitrariness of choosing diffusion matrix $D(x)$, while our results suggest that even if $D(x)$ is fixed, the decomposition is also not unique. 

Our results on the non-uniqueness of SDE decomposition raise a meaningful question both experimentally and theoretically: given the system described by (\ref{eqn:original}), how would nature choose one particular process with decomposition form (\ref{eqn:transform}) from all the candidates?  Could there exist any other restrictions on $S$ and $A$ besides condition (\ref{eqn:condition}) which helps determine the decomposed process uniquely?

\section*{Acknowledgement}

T. Li acknowledges the support of NSFC under grants 11171009, 11421101, 91130005 and the National Science Foundation for Excellent Young Scholars (Grant No. 11222114). The authors are grateful to Hong Qian, Hao Ge and Xiaoguang Li for helpful discussions. We also thank Ying Tang for introducing the reference \cite{ma2015complete}.

\bibliographystyle{ieeetr}
\bibliography{reference}

\newpage
\renewcommand{\theequation}{S-\arabic{equation}}
\setcounter{equation}{0} 

\begin{center}
{\bf \Large APPENDIX}
\end{center}

\begin{appendix}	
\section{Laplace's Method and Laplace Principle}

Laplace's method is used for approximating integrals of exponential type, stating that 
\begin{equation}\label{eqn:lapmeth}
\int_{a}^{b} e^{g(x)/\varepsilon}\sim \sqrt{\frac{2\pi\varepsilon}{|g''(x_{0})|}}e^{g(x_{0})/\varepsilon}, \quad \text{as $\varepsilon\to 0+$}
\end{equation}
where $x_{0}$ is assumed to be the unique maximum of $g(x)$.

In large deviation theory, the logarithmic form of Eq.  (\ref{eqn:lapmeth}) is commonly used, known as the Laplace principle. Suppose $A$ is a regular subset (Borel set in mathematics) and $\varphi$ is a measurable function, then 
\begin{equation*}
\lim_{\varepsilon\to 0}\varepsilon\ln\int_{A}e^{-\varphi(x)/\varepsilon} dx=-\inf_{x\in A}\varphi(x).
\end{equation*}
In the main text, we also formally adopt the Laplace principle in the infinite dimensional path space.

\section{Construction of Local Quasi-Potential}

We will obtain the expression of local quasi-potential for the 1-D example appeared in Section \ref{sec:localqp}. The SDE under consideration is  
\begin{equation*}
dX_{t}=-\nabla V(X_{t})dt+\sqrt{2\varepsilon}dW_{t},
\end{equation*}
where the potential $V(x)$ has two local minimum points $x_{1}$ and $x_{2}$ ($x_{1}<x_{2}$ and $V(x_{1})<V(x_{2})$) and one local maximum point $x_{3}$ in-between. The local quasi-potential with respect to stable point $x_{1}$ is defined by 
\begin{equation*}
\phi^{QP}_{loc}(x;x_{1})=\inf\limits_{T>0}\inf\limits_{\psi(0)=x_{1},\psi(T)=x}S_{T}[\psi]
\end{equation*}
where $S_{T}[\psi]$ is given in \eqref{eq:SLg}. The value of $\phi^{QP}_{loc}(x;x_{1})$ can be computed depending on the different locations of $x$. 
\begin{description}
\item[Case 1:] $x\leq x_{3}$. We have
\begin{equation*}
\begin{split}
S_{T}[\psi] &=\frac{1}{4}\int_{0}^{T}|\dot{\psi}-\nabla V(\psi)|^{2}dt+\int_{0}^{T}\dot{\psi}\cdot \nabla V(\psi) dt \\
                  &\geq \int_{0}^{T}\dot{\psi}\cdot\nabla V(\psi) dt =V(\psi(T))-V(\psi(0)) \\
                  &=V(x)-V(x_{1}).
\end{split}
\end{equation*}
The equality holds when we take $\hat{\psi}$ such that $\dot{\hat{\psi}}=\nabla V(\hat{\psi})$ and $\hat{\psi}(0)=x_{1},\hat{\psi}(T)=x,T=+\infty$, which corresponds to the steepest ascent path connecting $x_{1}$ and $x$. Therefore $\phi^{QP}_{loc}(x;x_{1})=V(x)-V(x_{1})$, $x\leq x_{3}$.

\item[Case 2:] $x_{3}<x<x_{2}$.

In this case, the path connecting $x_{1}$ and $x$ passes the point $x_{3}$ and we assume that $\psi(T_{1})=x_{3}$. We have
\begin{equation*}
\begin{split}
S_{T}[\psi] &=\frac{1}{4}\int_{0}^{T_{1}}|\dot{\psi}+\nabla V(\psi)|^{2}dt+\frac{1}{4}\int_{T_{1}}^{T}|\dot{\psi}+\nabla V(\psi)|^{2}dt \\
                  &=\frac{1}{4}\int_{0}^{T_{1}}|\dot{\psi}-\nabla V(\psi)|^{2}dt+\int_{0}^{T_{1}}\dot{\psi}\cdot \nabla V(\psi) dt +\frac{1}{4}\int_{T_{1}}^{T}|\dot{\psi}+\nabla V(\psi)|^{2}dt\\
                  &\geq \int_{0}^{T_{1}}\dot{\psi}\cdot \nabla V(\psi) dt =V(\psi(T_{1}))-V(\psi(0)) \\
                  &=V(x_{3})-V(x_{1})
\end{split}
\end{equation*}
with the minimum action path satisfying $\dot{\hat{\psi}}=\nabla V(\hat{\psi})$ for $\hat{\psi}(t)<x_{3}$ and  $\dot{\hat{\psi}}=-\nabla V(\hat{\psi})$ for $x_{3}<\hat{\psi}(t)<x_{2}$, containing the steepest ascent path from $x_{1}$ to $x_{3}$ and the steepest descent path (the ODE path) from $x_{3}$ to $x$. Hence $\phi^{QP}_{loc}(x;x_{1})=V(x_{3})-V(x_{1}), x\leq x_{3}<x<x_{2}$.

\item[Case 3:] $x\geq x_{2}$.

We assume that $\psi(T_{1})=x_{3},\psi(T_{2})=x_{2}$. Then
\begin{equation*}
\begin{split}
S_{T}[\psi] &=\frac{1}{4}\int_{0}^{T_{1}}|\dot{\psi}+\nabla V(\psi)|^{2}dt+\frac{1}{4}\int_{T_{1}}^{T_{2}}|\dot{\psi}+\nabla V(\psi)|^{2}dt+\frac{1}{4}\int_{T_{2}}^{T}|\dot{\psi}+\nabla V(\psi)|^{2}dt \\
                  &=\frac{1}{4}\int_{0}^{T_{1}}|\dot{\psi}-\nabla V(\psi)|^{2}dt+\int_{0}^{T_{1}}\langle\dot{\psi}, \nabla V(\psi)\rangle dt \\
                   &+\frac{1}{4}\int_{T_{1}}^{T_{2}}|\dot{\psi}+\nabla V(\psi)|^{2}dt+\frac{1}{4}\int_{T_{2}}^{T}|\dot{\psi}-\nabla V(\psi)|^{2}dt+\int_{T_{2}}^{T}\dot{\psi}\cdot\nabla V(\psi) dt\\
                  &\geq \int_{0}^{T_{1}}\dot{\psi}\cdot\nabla V(\psi) dt +\int_{T_{2}}^{T}\dot{\psi} \cdot\nabla V(\psi) dt\\               
                  &=V(\psi(T_{1}))-V(\psi(0))+ V(\psi(T))-V(\psi(T_{2})) \\
                  &=V(x)-V(x_{2})+V(x_{3})-V(x_{1})
\end{split}
\end{equation*}
with the minimum action path satisfying $\dot{\hat{\psi}}=\nabla V(\hat{\psi})$ for $\hat{\psi}(t)<x_{3}$ or $\hat{\psi}(t)>x_{2},$ and $\dot{\hat{\psi}}=-\nabla V(\hat{\psi})$ for $x_{3}<\hat{\psi}(t)<x_{2}$. The path contains two ascent parts from $x_{1}$ to $x_{3}$ and $x_{2}$ to $x$, and a descent part from $x_{3}$ to  $x$. Hence $\phi^{QP}_{loc}(x;x_{1})=V(x)-V(x_{2})+V(x_{3})-V(x_{1}), x\geq x_{2}$.
\end{description}

The local quasi-potential with respect to $x_{2}$ can be calculated similarly.

\section{Proof of the Theorems}

We will prove Theorem \ref{thm:main} by constructing the desired $S$ and $A$ in the theorem, and the results in Theorem \ref{thm:sub} will be revealed during the construction process.

Recall the main idea of reconstruction procedure for SDE decomposition from the solution $\phi(x)$ of Hamilton-Jacobi  equation \eqref{eqn:hje}. If the the described procedure works, then Theorem \ref{thm:main} will be verified. In theoretical aspects one needs to ensure 
\begin{itemize}
\item The existence of solutions for the linear system arises in Step 2, which guarantees the existence of $G(x)$.

\item The invertibility of the constructed $G(x)$, which guarantees the existence of $S(x)$ and $A(x)$ in Step 3.

\item The verification of condition (\ref{eqn:condition}) for $S(x)$ and $A(x)$ obtained in Step 3, which guarantees the constructed $S,A$ and $\phi(x)$ are the desired quantities in Eq.  (\ref{eqn:transform}).
\end{itemize}

We will show that under the assumptions stated in Theorem \ref{thm:main},  all the  requirements above can be satisfied.  Theorems \ref{thm:main} and \ref{thm:sub} can then be proved as the result.

\subsection{The Existence of $G(x)$}\label{sec:existence}

From $G(x)+G^{T}(x)=2D(x)$ we can write $G(x)=D(x)+Q(x)$, where $Q(x)$ is an anti-symmetric matrix. Hence the existence of  $Q(x)$ such that 
\begin{equation}
Q(x)\nabla\phi(x)=-b(x)-D(x)\nabla\phi(x)
\label{eqn:Q}
\end{equation}
will imply the existence of $G(x)$ in Step 2 of the reconstruction procedure. We use the vector $q(x)=(q_{1}(x), q_{2}(x),\dots ,q_{n(n-1)/2}(x))^{T}$ to represent $Q(x)$ by
\begin{equation*}
Q(x)=\left(
\begin{matrix}
0 & q_{1} & q_{2} & \dots & q_{n-2} & q_{n-1} \\
-q_{1} & 0 & q_{n} & \dots & q_{2n-4} & q_{2n-3} \\
-q_{2} & -q_{n} & 0 & \dots & q_{3n-7} & q_{3n-6} \\
\vdots & \vdots & \vdots & \ddots & \vdots & \vdots \\
-q_{n-2} & -q_{2n-4} & -q_{3n-7} & \dots & 0 & q_{n(n-1)/2} \\
-q_{n-1} & -q_{2n-3} & -q_{3n-6} & \dots  & -q_{n(n-1)/2}  & 0\\
\end{matrix}
\right).
\end{equation*}
We can transform Eq.  (\ref{eqn:Q}) into the following linear system for vector $q(x)$:
\begin{equation}
\Psi(x) q(x)= -b(x)-D(x)\nabla\phi(x).
\label{eqn:linearsystem}
\end{equation}
The coefficient matrix $\Psi(x)$ has the form
\begin{equation*}
\Psi(x)=(\Psi_{1}(x),\Psi_{2}(x),\dots,\Psi_{n-1}(x)),
\end{equation*}
where the $i$-th block $\Psi_{i}(x)$ is an $n\times (n-i)$ matrix with the structure
\begin{equation*}
\Psi_{i}(x)=\left(
\begin{matrix}
O \\
\tilde{\Psi}_{i}(x)
\end{matrix}
\right), 
\end{equation*}
in which $O$ represents the $(i-1)\times(n-i)$ zero matrix and the structure of $\tilde{\Psi}_{i}(x)$ is 
\begin{equation*}
\tilde{\Psi}_{i}(x)=\left(
\begin{matrix}
\phi_{x_{i+1}} & \phi_{x_{i+2}} &  \phi_{x_{i+3}} & \dots & \phi_{x_{n-1}} &\phi_{x_{n}} \\
-\phi_{x_{i}} & 0 &0 & \dots & 0 & 0 \\
0 & -\phi_{x_{i}} & 0 &\dots & 0  & 0\\
0 & 0 & -\phi_{x_{i}} &\dots & 0 & 0\\
\vdots & \vdots & \vdots & \ddots  & \vdots  & \vdots\\   
0 & 0 & 0 & \dots & -\phi_{x_{i}} & 0 \\  
0 & 0 & 0 & \dots & 0 & -\phi_{x_{i}} \\  
\end{matrix}
\right)\in \mathbb{R}^{(n-i+1)\times(n-i)},
\end{equation*}
where $\phi_{x_{i}}$ denotes $\partial_{x_{i}}\phi(x)$. The concrete expression of $\Psi$ for dimension $n=4$ takes the following form
\begin{equation}
\Psi=\left(
\begin{matrix}
 \phi_{x_{2}} & \phi_{x_{3}}  &  \phi_{x_{4}} & 0 & 0& 0 \\
- \phi_{x_{1}} & 0 & 0 &  \phi_{x_{3}} & \phi_{x_{4}} & 0 \\
0 & - \phi_{x_{1}} & 0& - \phi_{x_{2}} & 0 &  \phi_{x_{4}} \\
0 &	0	&	 - \phi_{x_{1}}&	0 &  -\phi_{x_{2}} &  -\phi_{x_{3}}
\end{matrix}
\right).
\label{eqn:matrix4d}
\end{equation}

Based on the above manipulations, the existence of $G(x)$ is converted to the solvability of linear system (\ref{eqn:linearsystem}). This can be ensured by studying the following two cases.
\begin{itemize}
\item Case 1:  $\nabla\phi(x)\not=0$.

Notice that all column vectors of matrix $\Psi$ are orthogonal to $\nabla\phi$ and from Hamilton-Jacobi equation we know the right hand side $-b(x)-D\nabla\phi(x)$ is also orthogonal to $\nabla\phi$. Thus the column space of augmented matrix $A=(\Psi, -b-D\nabla\phi)$ is orthogonal to the non-zero vector $\nabla\phi$. This indicates that the column space cannot be the whole space $\mathbb{R}^{n}$ (otherwise $\nabla\phi=0$), so we have $\text{rank}(A)\leq n-1$. On the other hand, from $\nabla\phi\not=0$ we may assume $\phi_{x_{i_{0}}}\not=0$. Then there exists an $(n-1)\times(n-1)$ nonsingular diagonal sub-matrix of $\Psi$ with diagonal elements $\pm\phi_{x_{i_{0}}}$. Hence $n-1\leq \text{rank}(\Psi)\leq \text{rank}(A)\leq n-1$, which yields $\text{rank}(A)=\text{rank}(\Psi)=n-1$ and therefore guarantees the existence of solution $q(x)$.

\item Case 2: $\nabla\phi(x)=0$.
	
From the assumption that $b$ and $\nabla\phi$ have the same zeros, we must have $b(x)=0$. Then in Eq. (\ref{eqn:linearsystem}), the $\Psi$ on the left hand side is a zero matrix and the right hand side is a zero vector, therefore any $q(x)\in\mathbb{R}^{n}$ solves the linear system.
\end{itemize}

Hence we conclude that under the assumptions stated in Theorem \ref{thm:main}, the solution $q(x)$ of linear system (\ref{eqn:linearsystem}) always exists. This ensures the existence of $Q(x)$ and $G(x)$ in Step 2 of the reconstruction procedure.

\subsection{The Invertibility of $G(x)$}

To show that the matrix $G(x)$ constructed is invertible for any given $x$, we need to utilize the relation $G(x)=D(x)+Q(x)$ and the fact that $D(x)$ is positive definite.  
Assume that $y\in\mathbb{R}^{n}$ is the solution of linear system $G(x)y=0$. We then have 
\begin{equation*}
 0=y^{t}G(x)y=y^{t}[D(x)+Q(x)]y=y^{t}D(x)y.
\end{equation*}
From the positive definiteness of $D(x)$, we conclude that $y=0$. This ensures the invertability of $G(x)$. 

\subsection{Verification of Conditions for $S(x)$ and $A(x)$}

With the constructed $G(x)$, we define 
$$S(x)=\frac{1}{2}[G^{-1}(x)+G^{-T}(x)]\quad \text{and}\quad   A(x)=\frac{1}{2}[G^{-1}(x)-G^{-T}(x)].$$ 
Direct calculation shows that
\begin{align*}
[S(x)+A(x)]b(x)& =G^{-1}(x)b(x)=-\nabla\phi(x), \\
[S(x)+A(x)]D(x)[S(x)-A(x)] &=G^{-1}(x)\frac{1}{2}[G(x)+G^{T}(x)]G^{-T}(x) \notag\\
 &=\frac{1}{2}[G^{-T}(x)+G^{-1}(x)]=S(x),
\end{align*}
which concludes the proof of Theorem \ref{thm:main}. The results also indicate that the constructed $S(x)$ and $A(x)$ from the procedure satisfies the condition (\ref{eqn:condition}) and thus $S,A,\phi$ are the desired quantities in the transformed stochastic process (\ref{eqn:transform}). 

Moreover, the argument in Case 1 of Appendix \ref{sec:existence} also implies that the degrees of freedom for solutions of linear system (\ref{eqn:linearsystem}) is $n(n-1)/2-(n-1)=(n-1)(n-2)/2$ provided that $\nabla\phi(x)\not=0$. Since $G(x)$ has the structure
\begin{equation*}
G(x)=G^{*}(x)+\sum_{k=1}^{(n-1)(n-2)/2} \lambda_{k}(x) Q_{k}(x),
\end{equation*}
where $G^{*}(x)$ is a special solution and $Q_{k}(x)$ a set of linearly independent fundamental solutions, then $G^{-1}(x)$ and the constructed $S(x)$ and $A(x)$ also possess the degrees of freedom $(n-1)(n-2)/2$, which leads to the conclusion of Theorem \ref{thm:sub}.

\section{Global Representation of $G, S, A$}\label{sec:remark}

In previous discussions, we have shown the local existence of the matrix $G(x)$ pointwisely under certain conditions.  Moreover,  if $p_{0}$ is the non-singular point of $\phi(x)$, we can choose this local $G(p_{0})$ with degrees of freedom $(n-1)(n-2)/2$ , and for singular points, any matrix $G(p_{0})$ with symmetric part $D(p_{0})$ applies. 

However, it is not sufficient to represent the structure of global mapping $G(x)$ by simply assigning it to a candidate $G(x_{0})$ for every point $x_{0}$ in the domain, because the A-type Fokker-Planck equation (\ref{eqn:aint}) also imposes additional requirements on the differentiability of  global mapping $G(x)$. Below we suggest a possible tactic to represent a family of feasible global $G(x)$ (and therefore $S(x)$ and $A(x)$) starting from a given special differentiable solution.

For simplicity, we first consider the example of $n=4$. Suppose that we have already obtained one special solution $q_{0}(x)$ for the linear system (\ref{eqn:linearsystem}) which is differentiable. Then solving the linear system at $p_{0}$ which is a non-singular point of $\phi(x)$  (without loss of generality, assume that $\phi_{x_{1}}(p_{0})\not=0$), we get the following representation of solution to Eq. (\ref{eqn:linearsystem}): 
\begin{equation}
q(p_{0})=q_{0}(p_{0})+k_{1}q_{1}(p_{0})+k_{2}q_{2}(p_{0})+k_{3}q_{3}(p_{0}),  \quad k_{1},k_{2},k_{3} \in\mathbb{R}
\label{eqn:represent}
\end{equation}
where $q_1=(\phi_{x_{3}},-\phi_{x_{2}},0,\phi_{x_{1}},0,0)^{T}$, $q_{2}=(\phi_{x_{4}},0,-\phi_{x_{2}},0,\phi_{x_{1}},0)^{T}$,  $q_{3}=(0,\phi_{x_{4}},-\phi_{x_{3}},0,0,\phi_{x_{1}})^{T}$. 

Note that although the $q_{i}$ in Eq. (\ref{eqn:represent}) only takes value at $p_{0}$, they are acturally defined in the whole domain. So we can define the global anti-symmetric matrix $Q_{\lambda}(x)$ by
\begin{equation*}
Q_{\lambda}(x)=Q_{0}(x)+\lambda_{1}(x)Q_{1}(x)+\lambda_{2}(x)Q_{2}(x)+\lambda_{3}(x)Q_{3}(x)
\end{equation*}
where $Q_{i}$ are the corresponding anti-symmetric matrix for vector $q_{i}$ defined through (\ref{eqn:Q}) and $\lambda_{i}(x)$ are arbitrary smooth functions of $x$.
Specifically we have
\begin{equation*}
Q_{1}=\left(
\begin{matrix}
0 & \phi_{x_{3}} & -\phi_{x_{2}} & 0  \\
-\phi_{x_{3}} & 0 & \phi_{x_{1}} & 0  \\
\phi_{x_{2}} & -\phi_{x_{1}} & 0 & 0  \\
0 & 0 & 0& 0 \\
\end{matrix}
\right),
\end{equation*}
\begin{equation*}
Q_{2}=\left(
\begin{matrix}
0 & \phi_{x_{4}} & 0 & -\phi_{x_{2}}  \\
-\phi_{x_{4}} & 0 & 0 & \phi_{x_{1}}   \\
0 & 0 & 0& 0 \\
\phi_{x_{2}} & -\phi_{x_{1}} & 0 & 0  \\
\end{matrix}
\right),
\end{equation*}
and
\begin{equation*}
Q_{3}=\left(
\begin{matrix}
0 & 0 & \phi_{x_{4}} & -\phi_{x_{3}}   \\
0 & 0 & 0& 0 \\
-\phi_{x_{4}} & 0 & 0 & \phi_{x_{1}}   \\
\phi_{x_{3}} & 0 & -\phi_{x_{1}} & 0   \\
\end{matrix}
\right).
\end{equation*}
The global $G_{\lambda}(x),S_{\lambda}(x)$ and $A_{\lambda}(x)$ can be subsequently constructed from $Q_{\lambda}(x)$, which are all differentiable as long as $\phi(x)$ is sufficiently smooth. Similar argument works for any dimension $n\ge 3$. 

The arguments above are dependent on the existence of a specific differentiable $Q(x)$.  In fact,  we can find an explicit solution of anti-symmetric matrix $Q_{0}(x)$ to Eq. (\ref{eqn:Q})
\begin{equation}
Q_{0}(x)=\frac{\nabla\phi(b+D\nabla\phi)^{t}-(b+D\nabla\phi)\nabla^{t}\phi}{\nabla^{t}\phi\nabla\phi}.
\end{equation} 
This can be directly verified by using the orthogonality between $\nabla\phi$ and ($b+D\nabla\phi$). As long as some smoothness conditions of $b(x)$ and $\phi(x)$ can be guaranteed, the proposed $Q_{0}(x)$ is differentiable and therefore may serve as the starting point of the tactics described above.
\end{appendix}
\end{document}